\documentclass[sn-mathphys,Numbered]{sn-jnl}

\usepackage{graphicx}%
\usepackage{multirow}%
\usepackage{amsmath,amssymb,amsfonts}%
\usepackage{amsthm}%
\usepackage{mathrsfs}%
\usepackage[title]{appendix}%
\usepackage{xcolor}%
\usepackage{textcomp}%
\usepackage{manyfoot}%
\usepackage{booktabs}%
\usepackage{algorithm}%
\usepackage{algorithmicx}%
\usepackage{algpseudocode}%
\usepackage{listings}%
\usepackage{lineno}

\theoremstyle{thmstyleone}%
\theoremstyle{thmstyletwo}%

\theoremstyle{thmstylethree}%

\begin{document}

\title[AMOC-TEMP]{Failure to track a stable AMOC state under rapid climate change} 


\author*[1]{\fnm{Ren\'e M.} \spfx{van} \sur{Westen}}\email{r.m.vanwesten@uu.nl}

\author[1]{\fnm{Reyk} \sur{B\"orner}}\email{r.borner@uu.nl}

\author[1]{\fnm{Henk A.} \sur{Dijkstra}}\email{h.a.dijkstra@uu.nl}

\affil[1]{\orgdiv{Department of Physics}, \orgname{Institute for Marine and Atmospheric research Utrecht, Utrecht University}, \orgaddress{\street{Princetonplein 5}, \city{Utrecht}, \postcode{3584 CC},  \country{the Netherlands}}}


\abstract{{\bf The Atlantic Meridional Overturning Circulation (AMOC) is a tipping element of the climate system. The 
current  estimate of the global warming threshold for the onset of an  AMOC collapse is +4.0$^{\circ}$C (uncertainty range 1.4--8$^\circ$C). 
However,  such a threshold may not be meaningful because AMOC stability rather depends on the rate of radiative forcing change.
Here, we identify an AMOC stabilising  mechanism  that operates on timescales longer than present-day radiative forcing increase.
 Slow forcing permits coherent adjustment of surface and interior ocean 
properties, supported by enhanced evaporation and reduced sea-ice extent, counteracting destabilising feedbacks.  
We explicitly demonstrate this mechanism in a slow CO$_2$ ramp (+0.5~ppm~yr$^{-1}$) climate model simulation, in  
which the AMOC remains stable  up to +5.5$^{\circ}$C of global warming. 
By contrast, under faster CO$_2$ ramps, the AMOC collapses at substantially lower warming levels (+2$^{\circ}$C).
Our findings demonstrate rate-induced AMOC tipping and imply that 
limiting the rate of greenhouse gas emissions is critical for reducing the near-term risk of an AMOC collapse. 
} 
}

\keywords{AMOC, Climate change, Tipping risk, Rate-induced tipping}



\maketitle

\section*{Main}

Earth’s climate system contains various tipping elements in the biosphere, cryosphere and hydrosphere \cite{Lenton2025}. 
These tipping elements may undergo a transition upon exceeding a critical forcing, often expressed as global warming thresholds \cite{Armstrong2022}. 
For the Atlantic Meridional Overturning Circulation (AMOC), the threshold for a collapse is estimated to be at
global warming levels of +4$^{\circ}$C, with a range from +1.4$^{\circ}$C to +8.0$^{\circ}$C \cite{Armstrong2022}. 
In particular the lower bound of such estimates is important for society to identify safe operating spaces that minimise the risk of disruptive climate change \cite{Wunderling2023,Moller2024}.

However, estimating a critical warming threshold for the AMOC poses several challenges, 
from limited paleo-climatic evidence and simulated tipping events to model biases causing large uncertainty in the AMOC response to radiative forcing  \cite{Rahmstorf2005,Kriegler2009,Armstrong2022,vanWesten2025a}. 
A key underlying assumption behind such a threshold is that the present-day AMOC state will lose stability if the corresponding warming level is exceeded, triggering the onset of a collapse \cite{Lenton2008,Armstrong2022,Lenton2025}. 
Combining this assumption with the future warming levels projected by models forced with the Shared Socioeconomic Pathways (SSPs), 
it seems possible to determine when an AMOC collapse would begin.

Yet, many results indicate that the evolution of the
AMOC depends on the rate of radiative forcing change -- in observations \cite{Dima2025}, conceptual climate models \cite{Stocker1997}, 
early global climate models \cite{Manabe1999}, and recently in a low-resolution earth system model \cite{Hankel2025}. 
In addition, modeling studies have found sustained strong AMOC states in very warm climates \cite{Bonan2022,Nobre2023,Curtis2024,Willeit2024b}, 
for example under doubled or quadrupled CO$_2$ levels. 
While studies have emphasised the forcing rate effects on the magnitude of AMOC weakening \cite{Manabe1999, Hankel2025}, 
the question whether the forcing rate alone can determine AMOC tipping has been less investigated. 
An early study, using a zonally-averaged three-basin ocean model \cite{Stocker1997, Schmittner1999}, demonstrated that a faster rate of radiative forcing can indeed collapse the AMOC while a slower forcing rate does not. 
This is one of the first examples showing rate-dependent tipping \cite{Ashwin2012,Wieczorek2023} of the AMOC, 
meaning that it fails to track its current stable state for sufficiently fast forcing change.

An additional problem with a warming threshold is that the main destabilising feedback mechanism of the AMOC may operate independently of global temperatures. 
While certain tipping elements (e.g., the Greenland Ice Sheet and West Antarctic Ice Sheet) are losing
stability directly due to higher temperatures \cite{Waibel2018,Mouginot2019,Boers2021,Turney2020}, 
the AMOC is destabilised through the salt-advection feedback \cite{Stommel1961,Rahmstorf1996,Marotzke2000,Vanderborght2025}. 
This feedback can be triggered by natural climate variability \cite{Castellana2019,Romanou2023,vanWesten2024c,Oh2025} 
and thermal or haline changes in surface buoyancy fluxes over the North Atlantic Ocean \cite{Jackson2023b,vanWesten2024a,vanWesten2025e,Drijfhout2025}.
Much progress has been made in assessing the stability properties of the AMOC with respect to varying surface freshwater forcing \cite{Weijer2019,Dijkstra2026b}, 
which in isolation requires unrealistically large freshwater inputs to collapse the present-day AMOC \cite{vanWesten2025a}.
However, the present-day AMOC is exposed to substantial changes in radiative forcing, impacting both freshwater and heat forcings.
This anthropogenic forcing is expected to substantially weaken the AMOC \cite{Weijer2020,Bonan2025,Dijkstra2026} and possibly trigger a collapse \cite{Liu2017,Romanou2023,Drijfhout2025}. Nonetheless, the interacting effects of changing heat and freshwater fluxes under global warming remain insufficiently understood.

Motivated by these issues with a global warming threshold for an AMOC collapse, we here
investigate the radiative forcing path dependence of AMOC tipping in the Community Earth
System Model (CESM, version 1.0.5) and several models participating in the Coupled
Model Intercomparison Project (CMIP) phase 6. 
Starting from a recent quasi-equilibrium freshwater flux simulation with the CESM, in which an AMOC collapse has been found \cite{vanWesten2024a}, 
we extend earlier work on rate-dependence in more detail \cite{Stocker1997, Schmittner1999} and beyond
AMOC weakening \cite{Manabe1999, Hankel2025}. 
We demonstrate that sufficiently fast forcing can cause an AMOC collapse without crossing a tipping threshold in CO$_2$ or warming levels, 
questioning the suitability of global warming thresholds for assessing AMOC tipping risk.

\section*{Stable AMOC under extreme climate change}

We will present an in-depth analysis of a slow CO$_2$ ramp simulation  (+0.5~ppm~yr$^{-1}$) with the CESM,
together with a comparison using faster CO$_2$ ramps (+2.5~ppm~yr$^{-1}$ and +5.0~ppm~yr$^{-1}$) and also
different Representative Concentration Pathway (RCP) scenarios; the results for the RCP scenarios  were already presented previously \cite{vanWesten2025e}.
All climate model simulations are branched from a quasi-equilibrium hosing simulation 
under constant pre-industrial (PI) radiative forcing conditions \cite{vanWesten2023b, vanWesten2024a}, which
we need to describe in more detail first. 

\begin{figure}[h]
\hspace{-2cm}
\includegraphics[width=1.3\columnwidth, trim = {0cm 0cm 0cm 0cm}, clip]{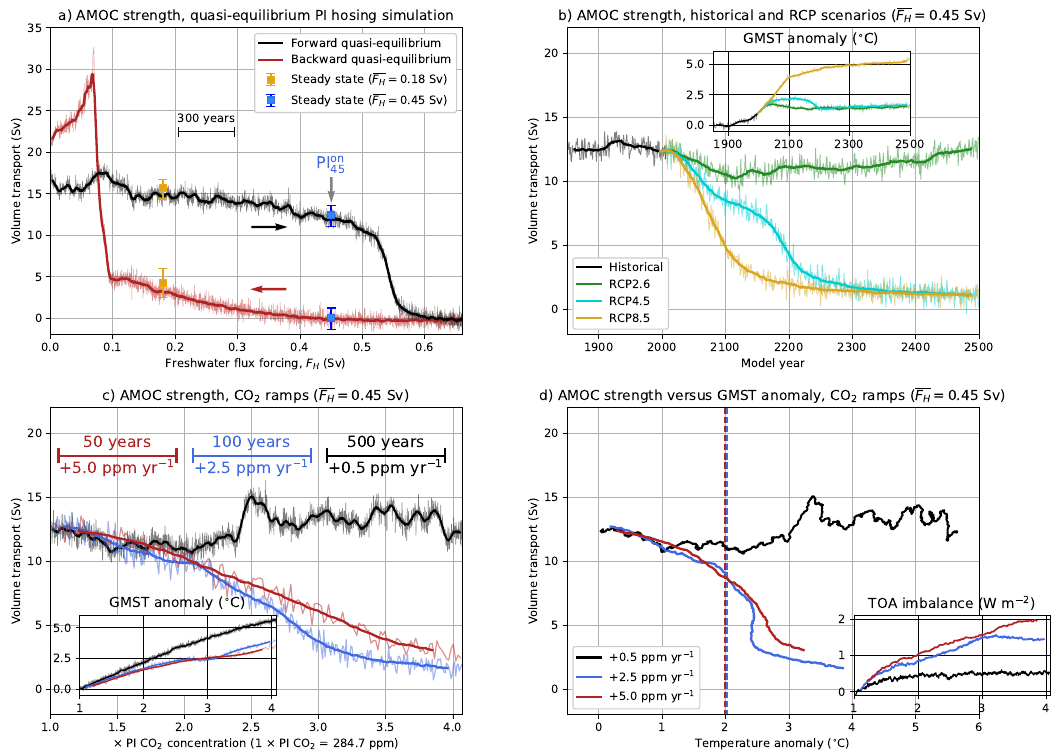} 
\caption{\textbf{Climate model simulations with the CESM.}
(a): The AMOC strength (at 26$^{\circ}$N and 1,000~m depth) for the quasi-equilibrium PI hosing simulation, including four statistical equilibria (last 50~model years).
(b): The AMOC strength for the historical and three extended RCP scenarios at fixed $\overline{F_H} = 0.45$~Sv. 
The inset shows the GMST anomaly compared to the historical period 1850--1899. 
(c \& d): The AMOC strength for the different CO$_2$ ramp simulations (c): against CO$_2$ concentration and (d): against GMST anomaly (compared to PI$_{45}^{\mathrm{on}}$), 
including the corresponding GMST values of AMOC collapse onset (Table~\ref{tab:Table_S1}).
The insets show (c) the GMST anomaly and (d) the TOA radiative imbalance, with horizontal axes in units of PI CO$_2$ concentration.
In all panels, the thin curves are yearly averages, whereas the thick curves are smoothed versions (25-year moving averages).}

\label{fig:Figure_1}
\end{figure}

In the quasi-equilibrium PI hosing simulation (Figure~\ref{fig:Figure_1}a), an additional freshwater flux ($F_H$) was gradually applied over the North Atlantic Ocean 
(20$^{\circ}$N -- 50$^{\circ}$N) and this forcing was compensated elsewhere (at the ocean surface) to conserve ocean salinity. 
Qualitatively comparable responses are found under small variations in hosing location \cite{Rahmstorf1996,Ma2024,Jackson2023b} 
and in the applied compensation protocol \cite{Rahmstorf1999, Dijkstra2024b,Mehling2026}, but both choices do influence quantitative thresholds.
The slowly-varying $F_H$ ensures that AMOC transitions are caused by intrinsic feedbacks. Multiple statistical equilibria (i.e., attracting steady states with time-invariant statistics) were obtained by branching off at selected values of $F_H$ and integrating for 500~years at fixed $F_H = \overline{F_H}$ \cite{vanWesten2024c,vanWesten2025a}.
The statistical equilibria have a radiative imbalance close to zero,
meaning that any remaining model drift is smaller than internal climate variability \cite{vanWesten2025c},
and are also shown in Figure~\ref{fig:Figure_1}a for $\overline{F_H} = 0.18$~Sv and  $\overline{F_H} = 0.45$~Sv.
In the following, we will use the PI steady state with a vigorous AMOC at $\overline{F_H} = 0.45$~Sv as a reference, 
which we refer to as PI$_{45}^{\mathrm{on}}$ (last~50~model years).

A tipping point (i.e., a saddle-node bifurcation) exists under an increasing freshwater flux forcing, which is located 
at $F_H \approx 0.5$~Sv for this CESM version \cite{vanWesten2025a}.
The existence of a tipping point in freshwater forcing is consistent with results across a hierarchy of 
climate models \cite{Stommel1961, Rahmstorf2005, Hawkins2011, Hu2012}.
Note that the AMOC tipping threshold is found at large $F_H$ values, about 65 times the present-day melt rate of the Greenland Ice Sheet \cite{vanWesten2024a},
which is related to persistent climate model biases \cite{vanWesten2024b,Dijkstra2024b,Boot2025}.
The real AMOC is likely closer to a tipping point \cite{vanWesten2025a,Dijkstra2026b}, and the imposed $\overline{F_H}$ should be regarded as a bias correction for AMOC stability. 
Here, we focus on PI$_{45}^{\mathrm{on}}$,
which lies in the destabilising regime of the salt-advection feedback and is hence more prone to AMOC transitions than at 
$\overline{F_H} = 0.18$~Sv \cite{vanWesten2025a}.  

Different radiative forcing simulations were branched from the end of PI$_{45}^{\mathrm{on}}$, keeping $\overline{F_H}$ fixed.
The historical forcing (1850 -- 2005) was followed by three RCP (2006 -- 2100) simulations. 
These RCP scenarios were subsequently continued to model year~2500 under their fixed greenhouse gas and aerosol concentrations of the year~2100 (Figure~\ref{fig:Figure_1}b).
This was done to determine statistical equilibria under different climate change scenarios \cite{vanWesten2025e}.
The AMOC recovers under RCP2.6, while the AMOC collapses under RCP4.5 and RCP8.5. 
Note that this does not guarantee that the AMOC will remain in a collapsed state beyond 2500 under these fixed climate change conditions  \cite{Bonan2022,Romanou2023,Nobre2023,Curtis2024}.
Consequently, it remains unclear whether the AMOC crosses a saddle-node bifurcation under radiative forcing increase, 
while strong evidence points to this being the case under increasing $F_H$ \cite{vanWesten2025a} (Figure~\ref{fig:Figure_1}a).
Hence, we avoid the term `tipping point' and instead refer to transitions to a weak alternative state as `AMOC collapse'. We determine the timing of the collapse onset using modern AMOC theory (see Methods and Table~\ref{tab:Table_S1}).
Even if the AMOC would eventually recover under fixed climate change conditions, 
the AMOC-induced impacts are considerable over the 300-year period (model year~2200 -- 2500) with persistently very weak AMOC strengths \cite{vanWesten2025c,vanWesten2025d}.

For RCP4.5 (RCP8.5), the collapse onset occurs around model year~2110 (2060), with an associated 
global mean near-surface temperature (GMST) anomaly of +2.19$^{\circ}$C (+2.78$^{\circ}$C) compared to the historical period of 1850 -- 1899 \cite{vanWesten2025e}.
The difference in the GMST value is 0.59$^{\circ}$C between RCP4.5 and RCP8.5,  highlighting a dependence on the radiative forcing agent (e.g., CO$_2$, methane, aerosols) pathways
that are relevant for AMOC responses \cite{Menary2020,Hassan2021,Hankel2025}.
Note that AMOC tipping under radiative forcing is also dependent on $\overline{F_H}$, as the AMOC recovers under RCP4.5 and $\overline{F_H} = 0.18$~Sv \cite{vanWesten2025e}; this simulation will also be analysed below. 

To further test the sensitivity of estimated warming thresholds for an AMOC collapse onset,
we performed three different CO$_2$ ramp simulations (again at fixed $\overline{F_H} = 0.45$~Sv),  where the CO$_2$ concentration increases from the PI level of 284.7~ppm (= $1\times$PI CO$_2$) at a certain rate. 
We choose a linear increase instead of the more common approach using exponential growth rates (e.g. \cite{Hankel2025}). 
This avoids very high CO$_2$ levels ($22\times$PI CO$_2$) and large CO$_2$ growth rates ($>100$~ppm~yr$^{-1}$) in our multi-centennial simulations.
Starting from the end of PI$_{45}^{\mathrm{on}}$, we linearly increased the CO$_2$ concentration up to $4.07\times$PI CO$_2$ (Figures~\ref{fig:Figure_1}c,d) with rates of
+0.5~ppm~yr$^{-1}$ (black curve, 1750~model years), +2.5~ppm~yr$^{-1}$ (blue curve, 350~model years), and +5.0~ppm~yr$^{-1}$ (red curve, 175~model years).
For reference, the current CO$_2$ rate of increase at the Mauna Loa Observatory is +2.6~ppm~yr$^{-1}$ (2015 -- 2025 average).

The AMOC collapses for both the +2.5~ppm~yr$^{-1}$ and +5.0~ppm~yr$^{-1}$ cases with an associated GMST anomaly of +2.02$^{\circ}$C and +1.99$^{\circ}$C 
compared to PI$_{45}^{\mathrm{on}}$, respectively (Table~\ref{tab:Table_S1}).
By contrast, in the slow +0.5~ppm~yr$^{-1}$ case the AMOC remains stable far above this apparent +2$^{\circ}$C threshold, 
reaching +5.5$^{\circ}$C over the last 50~model years.
This is an important indication that AMOC tipping under global warming can depend solely on the forcing rate, 
for which the trajectories fail to track a stable equilibrium under fast forcing changes.
This is corroborated by the radiative imbalance at the top of atmosphere (TOA, inset in Figure~\ref{fig:Figure_1}d), 
where TOA remains stable at +0.5~W~m$^{-2}$ for the +0.5~ppm~yr$^{-1}$ case, meaning that the climate system can reasonably track the shifting equilibrium under global warming.
For the faster ramps, the TOA imbalance increases throughout the simulations, indicative of a climate system that is getting further from equilibrium. 
Note that the warming level up to $2\times$PI CO$_2$ (inset in Figure~\ref{fig:Figure_1}c), for which all cases still exhibit a relatively strong AMOC, 
is larger for slower rates owing to the fact that the climate system has more time to warm, allowing slow amplifying climate feedbacks (e.g., sea-ice melt) to contribute.
The warming rates (up to $2\times$PI CO$_2$) are of course faster in +5.0~ppm~yr$^{-1}$ (+2.95$^{\circ}$C per century) than in +0.5~ppm~yr$^{-1}$ (+0.39$^{\circ}$C per century).

\section*{Density changes under slow warming}

In the following, we examine the slow CO$_2$ ramp simulation (+0.5~ppm~yr$^{-1}$) in greater detail to understand
why the AMOC remains stable under extreme global warming.
For convenience, we mostly report the slow CO$_2$ ramp in model years rather than in units of PI CO$_2$.

The AMOC in depth coordinates and meridional heat transport (MHT)  for model years~1 -- 50 ($\approx$ 1$\times$PI CO$_2$) and 1701 -- 1750 ($\approx$ 4$\times$PI CO$_2$) are shown in Figures~\ref{fig:Figure_S1}a,b, respectively. 
The overall overturning structures and MHTs do not change much,
though the depth of the northward overturning cell becomes smaller from about 2800~m depth (model years~1 -- 50) to about 2400~m depth (model years~1701 -- 1750).
The AMOC in density coordinates shifts to lighter water classes between the two periods (Figures~\ref{fig:Figure_S1}c,d), 
with the section-averaged depth levels of 20~m, 500~m and 3000~m getting lighter by about 0.90~kg~m$^{-3}$, 0.65~kg~m$^{-3}$ and 0.25~kg~m$^{-3}$, respectively.
The relatively light water masses near the surface are becoming lighter more rapidly than the relatively heavy water masses at depth, which is related to the greater warming closer to the surface (Figure~\ref{fig:Figure_S1}e).
The salinity  increases over most parts of the Atlantic Ocean (Figure~\ref{fig:Figure_S1}f), which makes water masses more dense and hence partially offsetting the warming-induced response of the density.  

\begin{figure}[h]

\begin{tabular}{c}

\includegraphics[width=1\columnwidth, trim = {0cm 0cm 0cm 0cm}, clip]{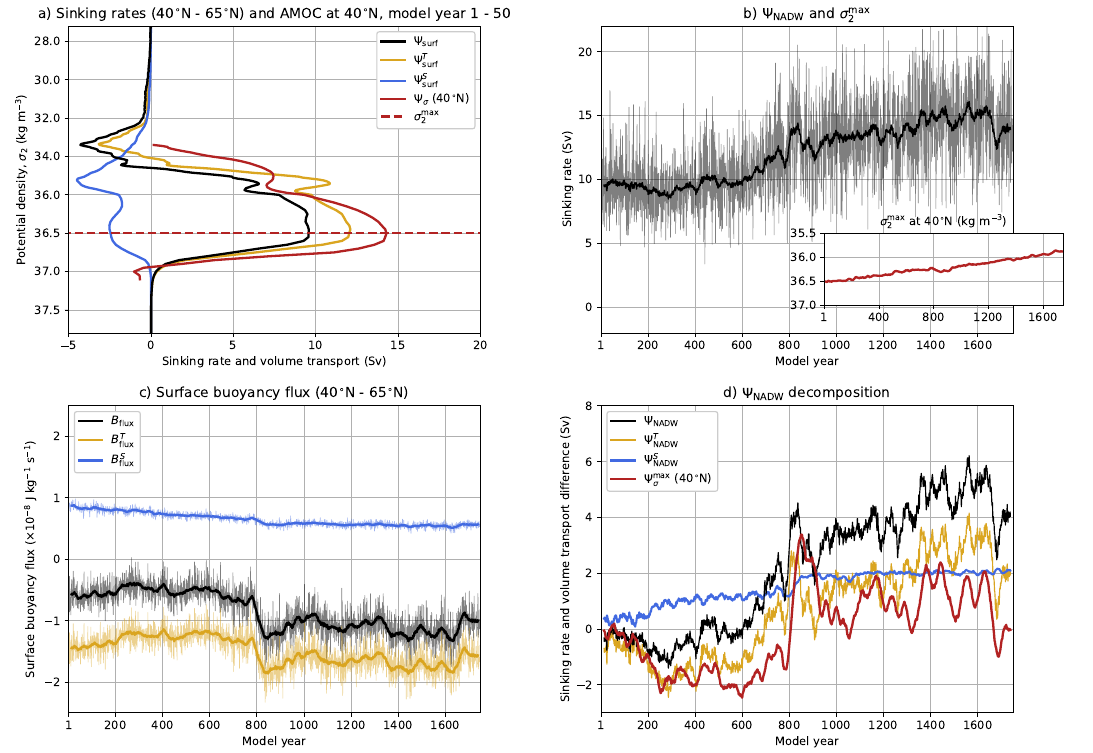} 
\end{tabular}

\caption{\textbf{Surface-forced AMOC in the slow CO$_2$ ramp} 
(a): The surface-forced AMOC between 40$^{\circ}$N and 65$^{\circ}$N ($\Psi_{\mathrm{surf}}$, sinking rates) over the first 50~model years,
including the temperature and salinity contributions.
The AMOC at 40$^{\circ}$N in density coordinates with its maximum indicated by the dashed red line ($\sigma_2^{\mathrm{max}}$) are also shown.
(b): The $\Psi_{\mathrm{NADW}}$ and $\sigma_2^{\mathrm{max}}$ (inset).
(c): The spatially-averaged (40$^{\circ}$N and 65$^{\circ}$N) surface buoyancy flux, decomposed into the heat and freshwater fluxes.
(d): The $\Psi_{\mathrm{NADW}}$ differences compared to PI$_{45}^{\mathrm{on}}$, including the temperature and salinity contributions.
The differences for the maximum AMOC strength at 40$^{\circ}$N are also shown.
The time series in panels~b,c,d are smoothed through a 25-year running mean (to reduce the variability).
}

\label{fig:Figure_2}

\end{figure}

Water mass transformation (WMT, see Methods) primarily takes place over the isopycnal outcropping region (40$^{\circ}$N -- 65$^{\circ}$N) in the North Atlantic Ocean, 
which is crucial for sustaining an adiabatic AMOC \cite{Nikurashin2012, Wolfe2014}.
Under the assumptions of thermal wind balance \cite{Vanderborght2025} and that WMT mainly takes place near the surface, 
the adiabatic AMOC can be reconstructed from surface buoyancy fluxes \cite{Walin1982, Marshall1999, Grist2009, Desbruyeres2019}. 
Indeed, the AMOC at 40$^{\circ}$N in density coordinates ($\Psi_{\sigma}$) and the surface-forced AMOC between 40$^{\circ}$N -- 65$^{\circ}$N ($\Psi_{\mathrm{surf}} = \Psi_{\mathrm{WMT}}(40^{\circ}$N$) - \Psi_{\mathrm{WMT}}(65^{\circ}$N)) 
closely resemble each other (red and black curve in Figure~\ref{fig:Figure_2}a, respectively).  
The surface-forced AMOC is mainly driven by oceanic heat loss to the atmosphere, whereas freshwater fluxes have a limited contribution (yellow and blue curves in Figure~\ref{fig:Figure_2}a), consistent with observations \cite{Marsh2000}. 
Surface-induced WMT rates that contribute to the water supply of the North Atlantic Deep Water (NADW) are found for density classes of $\sigma_2 \geq \sigma_2^{\mathrm{max}}$ \cite{vanWesten2025e},
with $\sigma_2^{\mathrm{max}}$ being the density level of the maximum AMOC strength at 40$^{\circ}$N (dashed red line in Figure~\ref{fig:Figure_2}a).
The reconstructed AMOC strength from surface buoyancy fluxes is then defined as $\Psi_{\mathrm{NADW}}(t) = \Psi_{\mathrm{surf}}(\sigma_2^{\mathrm{max}}(t))$,
which can also be decomposed into a thermal ($\Psi_{\mathrm{NADW}}^T$) and haline ($\Psi_{\mathrm{NADW}}^S$) contribution.
Note that $\Psi_{\mathrm{NADW}}$ approximates the actual AMOC strength (Figure~\ref{fig:Figure_2}d),  
because $\Psi_{\mathrm{NADW}}$ does not consider subsurface mixing and volume tendencies \cite{Groeskamp2019}.
In addition, WMT contributions outside the 40$^{\circ}$N -- 65$^{\circ}$N latitude band are relatively small \cite{vanWesten2025e}.

\begin{figure}[h!]

\hspace{-3cm}
\includegraphics[width=1.4\columnwidth, trim = {0cm 0cm 0cm 0cm}, clip]{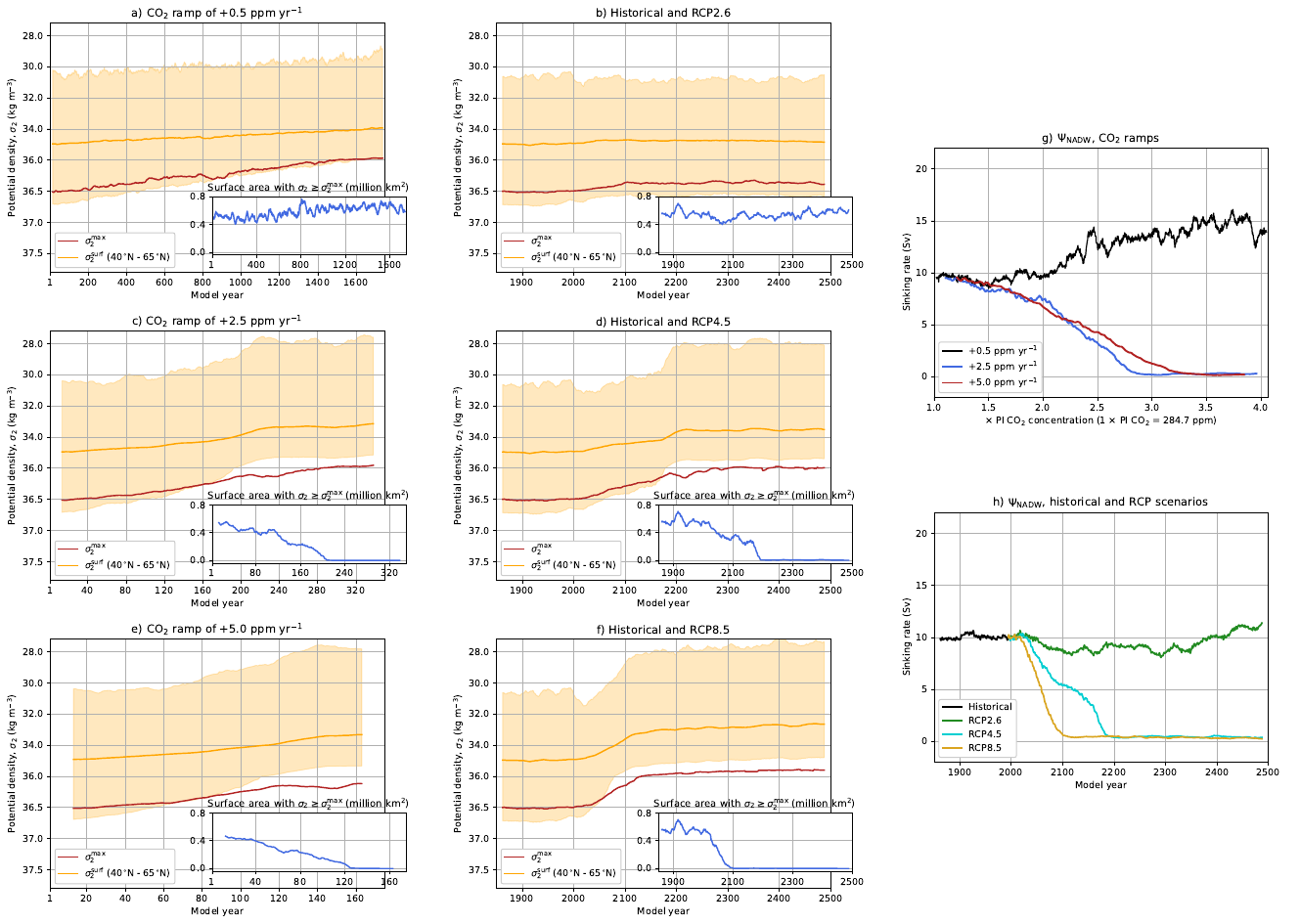} 

\caption{\textbf{Atlantic Ocean potential densities.}
(a -- f): The potential density in the North Atlantic Ocean for the surface ($\sigma_2^{\mathrm{surf}}$, 40$^{\circ}$N -- 65$^{\circ}$N, yellow curve) and at the maximum AMOC strength at 40$^{\circ}$N ($\sigma_2^{\mathrm{max}}$, red curve).
The yellow shading represents the variations in surface potential density, 
where for each month the lightest and heaviest potential densities (over 40$^{\circ}$N -- 65$^{\circ}$N) are retained
and are subsequently converted to yearly averages. 
The inset shows the surface area where the monthly surface potential density are heavier than $\sigma_2^{\mathrm{max}}$, which is subsequently converted to yearly averages.
The results are for the CO$_2$ ramps and historical and RCP scenarios, all have  $\overline{F_H} = 0.45$~Sv.
(g \& h): The $\Psi_{\mathrm{NADW}}$ for the CO$_2$ ramps (in units of PI CO$_2$) and historical and RCP scenarios.
All time series are smoothed through a 25-year running mean to reduce the variability.
}

\label{fig:Figure_3}
\end{figure}

The quantity $\Psi_{\mathrm{NADW}}$ is useful as near-zero values indicate that an adiabatic AMOC cannot be sustained  \cite{vanWesten2025e}.
In the slow CO$_2$ ramp, $\Psi_{\mathrm{NADW}}$ remains relatively large while $\sigma_2^{\mathrm{max}}$ decreases (Figure~\ref{fig:Figure_2}b).
This combined response means that both the interior and surface sinking region (40$^{\circ}$N -- 65$^{\circ}$N) are getting lighter at the same rate (Figure~\ref{fig:Figure_3}a),
such that adiabatic pathways remain open and support a strong AMOC. This also holds for the RCP2.6 simulation (Figure~\ref{fig:Figure_3}b).
The retreating sea-ice cover further activates additional sinking regions over the first 800~model years (inset in Figure~\ref{fig:Figure_3}a), 
which will be discussed in more detail below.
For the faster CO$_2$ ramps, as well as RCP4.5 and RCP8.5 simulations (Figure~\ref{fig:Figure_3}c--f),
the surface sinking region is getting lighter faster than $\sigma_2^{\mathrm{max}}$ and this closes the adiabatic AMOC pathways. 
For these cases, in which the AMOC collapses, near-zero values of $\Psi_{\mathrm{NADW}}$ (Figures~\ref{fig:Figure_3}g,h) happen about 100~years before the AMOC  strength is strongly reduced ($< 5$~Sv).
This  demonstrates that rate-dependent effects are important
not only for AMOC weakening \cite{Hankel2025}, but also for AMOC collapse scenarios.

The $\Psi_{\mathrm{NADW}}$ would, by construction, remain unchanged under a uniform shift in the ocean’s potential density,
meaning that $\Psi_{\mathrm{NADW}}$ variations are then linked to surface buoyancy fluxes changes ($B_{\mathrm{flux}}$, Figure~\ref{fig:Figure_2}c). In the slow CO$_2$ ramp simulation, the quantities $\Psi_{\mathrm{NADW}}$ and $B_{\mathrm{flux}}$ are closely related ($R^2 = 0.896$),
but note that a part of $B_{\mathrm{flux}}$ does not supply water to the adiabatic AMOC pathways.
Both surface heat fluxes ($B_{\mathrm{flux}}^T$) and surface freshwater fluxes ($B_{\mathrm{flux}}^S$) decline over time, 
resulting in larger $\Psi_{\mathrm{NADW}}^T$ and $\Psi_{\mathrm{NADW}}^S$, respectively (Figure~\ref{fig:Figure_2}d).
The initial AMOC weakening during the first 300~model years is attributed to changing heat fluxes, 
whereas freshwater fluxes have the opposite response.

The $B_{\mathrm{flux}}$ framework also allows identifying stabilising and destabilising factors.
For the surface heat fluxes in the slow CO$_2$ ramp (Figure~\ref{fig:Figure_S2}a), the dominant contributions come from more outgoing longwave radiation (stabilising) due to higher sea surface temperatures, more latent heat release (stabilising) due to enhanced evaporation rates, and
more shortwave absorption (destabilising) due to a lower albedo. 
The North Atlantic sea-ice extent retreats northward under higher temperatures (inset in Figure~\ref{fig:Figure_S2}b)
and this limits sea-ice insulation effects that tend to destabilise the AMOC \cite{Lin2023, vanWesten2024c}.
Another stabilising sea-ice effect is the reduced melting contribution to the surface freshwater flux (Figure~\ref{fig:Figure_S2}b). 
This sea ice forms around Greenland and by advection ends up in the 40$^{\circ}$N -- 65$^{\circ}$N latitude band, where it melts during boreal spring \cite{Vanderborght2025}.
Enhanced evaporation (stabilising) and precipitation (destabilising) rates also have a substantial contribution, but their opposing effects compensate to about zero.
The remaining $B_{\mathrm{flux}}$ components have a relatively small contribution. 

Almost all $B_{\mathrm{flux}}$ components have the opposite response in the faster CO$_2$ ramps (Figures~\ref{fig:Figure_S2}c,d,e,f) compared to the slow CO$_2$ ramp;
the faster CO$_2$ ramp responses are very similar to those under RCP4.5 and RCP8.5 (for the RCPs, see \cite{vanWesten2025e} or online material).
Recent work suggests that a sign change in $B_{\mathrm{flux}}$ (from negative to positive) can be used to indicate the onset of the AMOC collapse \cite{Willeit2024a,vanWesten2025e}, 
which is indeed the case for the faster CO$_2$ ramps, RCP4.5 and RCP8.5 (Figures~\ref{fig:Figure_S2}g,h).
By contrast, a more negative  $B_{\mathrm{flux}}$ indicates a relatively stable AMOC in the slow CO$_2$ ramp and RCP2.6 simulations.

In the slow CO$_2$ ramp, the AMOC strength undergoes a relatively rapid increase by about 3~Sv around model year~800 (Figure~\ref{fig:Figure_1}c), 
which  is related to the retreating sea-ice extent over the North Atlantic Ocean. 
Upon initialisation from PI$_{45}^{\mathrm{on}}$, the sea ice extends relatively far south (Figure~\ref{fig:Figure_S3}a), 
owing to a weak AMOC strength of about 13~Sv under the background hosing of $\overline{F_H} = 0.45$~Sv.
Sea ice limits ocean-atmosphere exchange and deep convection \cite{Lin2023}; hence the mixed layer depth is relatively shallow ($< 100$~m) where sea ice is present.
Importantly, there is initially no deep convection over the Labrador Sea due to the extensive sea-ice extent.
When sea ice retreats poleward under higher temperatures, the Labrador Sea becomes less sea-ice covered and this allows for deeper mixed layer depths (Figures~\ref{fig:Figure_S3}b,c,d).
Vertical mixing brings relatively warm subsurface water masses to the surface,  strongly contributing to the retreating sea-ice extent over the Labrador Sea, and giving rise to nonlinear responses under the linear CO$_2$ ramp.
Labrador Sea deep convection starts from model~year~770 and onwards, followed by the 3~Sv increase in AMOC strength around model year~800.
The role of deep convection on WMT rates is quite complex \cite{Bruggemann2019}, 
and this convection certainly plays an important role in modulating AMOC strength and collapsing the AMOC \cite{Drijfhout2025}. 
Hence, we expect that the 3~Sv increase can only be found when additional deep convection sites are activated as a result of retreating sea ice.

In summary, the analysis of the CO$_2$ ramp simulations and comparison with the RCP scenario runs (Figure~\ref{fig:Figure_3}) 
show that the ocean response to radiative forcing changes depends qualitatively on the forcing rate.
In the slow CO$_2$ ramp, in which the atmopsheric warming rate is (much) smaller than in the faster CO$_2$ ramps, RCP4.5 and RCP8.5, 
the AMOC remains stable up to +5.5$^{\circ}$C warming. 
In fact, from the $\Psi_{\mathrm{NADW}}$ and $B_{\mathrm{flux}}$ analysis, we conclude that the AMOC is getting more stable in warmer climates. 

\section*{Freshwater budget under climate change}

\begin{figure}[h!]

\hspace{-3cm}
\includegraphics[width=1.4\columnwidth, trim = {0cm 0cm 0cm 0cm}, clip]{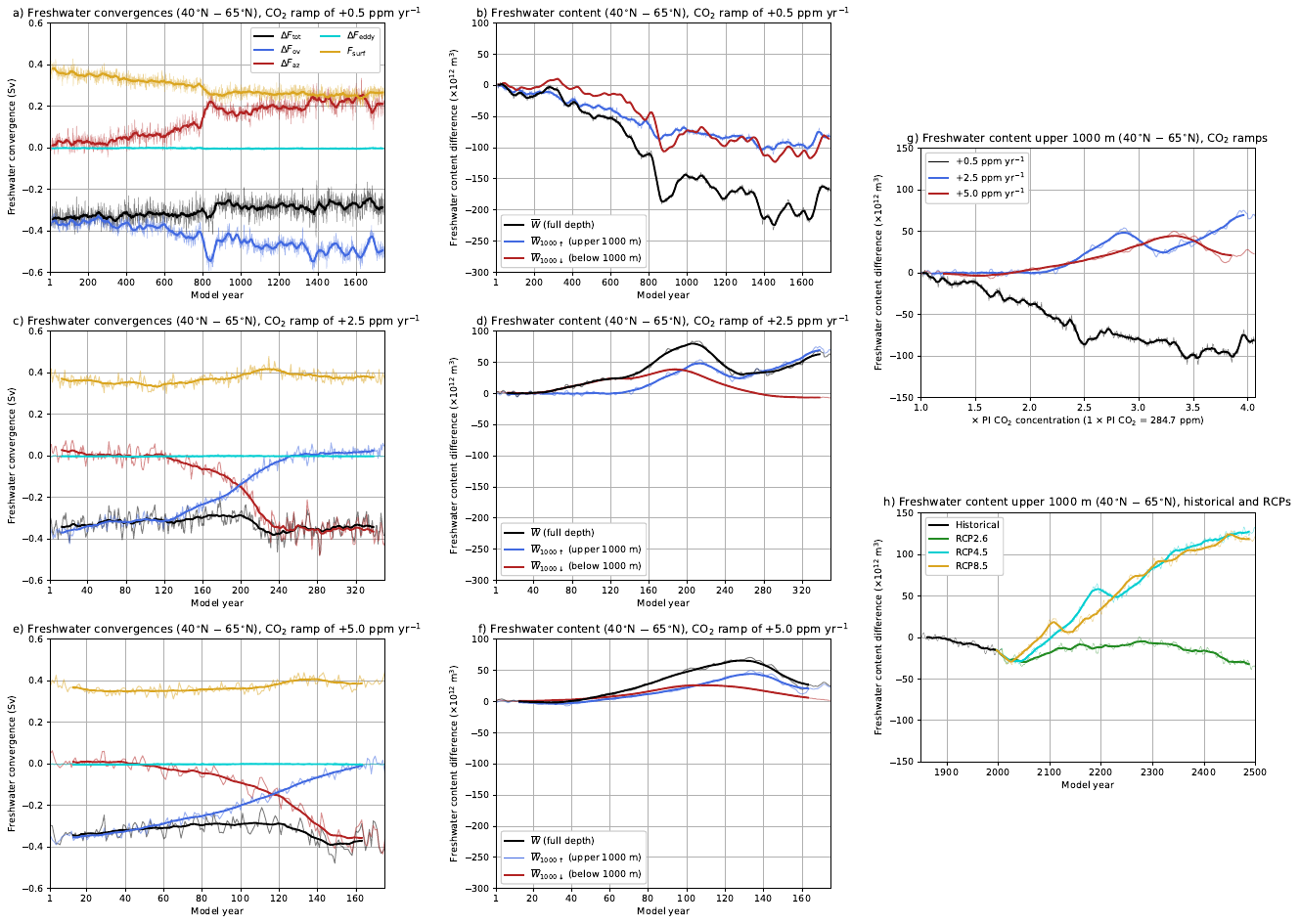} 

\caption{\textbf{Freshwater convergence over the North Atlantic Ocean.}
(Left column): The meridional freshwater convergences (40$^{\circ}$N -- 65$^{\circ}$N) for the different CO$_2$ ramps, including the surface freshwater flux (legend in panel~a). 
(Middle column): The freshwater content (40$^{\circ}$N -- 65$^{\circ}$N) difference (compared to PI$_{45}^{\mathrm{on}}$) for the different CO$_2$ ramps,
which is split into an upper 1,000~m and below 1,000~m contribution. 
(Right column): The upper 1,000~m freshwater content difference for the CO$_2$ ramps (compared to PI$_{45}^{\mathrm{on}}$, in units of PI CO$_2$) 
and historical and RCP scenarios (compared to the historical period of 1850 -- 1899).
In all panels, the thin curves are yearly averages, whereas the thick curves are smoothed versions (25-year moving averages).
}

\label{fig:Figure_4}
\end{figure}

Ultimately, an AMOC collapse is caused by net freshwater accumulation over the northern North Atlantic, 
and is driven by the destabilising salt-advection feedback  \cite{Stommel1961,Rahmstorf1996,Marotzke2000}.
This feedback can be activated by noise, freshwater perturbations or transient climate change \cite{Romanou2023,vanWesten2024a,Drijfhout2025,Oh2025}.
There is also a heat-advection feedback that tends to stabilise the AMOC \cite{Vanderborght2025}, 
but its effect is smaller than that of freshwater accumulation when expressed in terms of ocean buoyancy (\cite{vanWesten2025e} and Figures~\ref{fig:Figure_S4}a,c,e).
Hence, we focus on the freshwater convergence between 40$^{\circ}$N -- 65$^{\circ}$N of the Atlantic (Figure~\ref{fig:Figure_4}, see Methods).

For the slow CO$_2$ ramp simulation, the total freshwater convergence $\Delta F_{\mathrm{tot}}$ remains fairly constant over the first 800~model years and thereafter slightly increases (Figure~\ref{fig:Figure_4}a).
The total convergence is decomposed into its contributing factors, indicating
that the overturning convergence is declining ($\Delta F_{\mathrm{ov}}$, i.e., salinifying) while being balanced by the azonal (gyre) convergence ($\Delta F_{\mathrm{az}}$, i.e., freshening).
The declining $\Delta F_{\mathrm{ov}}$ values are consistent with the relatively large salinity increase south of 40$^{\circ}$N (Figure~\ref{fig:Figure_S1}f),
such that the AMOC imports net salinity into the 40$^{\circ}$N -- 65$^{\circ}$N latitude band.
However, these responses of $\Delta F_{\mathrm{ov}}$ and $\Delta F_{\mathrm{az}}$ cannot explain the lower freshwater content ($\overline{W}$, i.e., salinifying, Figure~\ref{fig:Figure_4}b) 
when considering the constant $\Delta F_{\mathrm{tot}}$ over the first 800~model years.
These declining values of $\overline{W}$ are primarily driven by a smaller freshwater input through the surface ($F_{\mathrm{surf}}$),
which was extensively discussed above (Figure~\ref{fig:Figure_S2}).

The freshwater budget over 40$^{\circ}$N -- 65$^{\circ}$N evolves completely differently for the faster CO$_2$ ramps (Figures~\ref{fig:Figure_4}c,d,e,f).
The overturning convergence is strongly increasing and, together with net freshwater accumulation over the upper 1,000~m, 
is indicative of the dominant and destabilising salt-advection feedback that causes the AMOC to collapse.
Similar to the slow CO$_2$ ramp,  $\Delta F_{\mathrm{az}}$ has the opposite response than $\Delta F_{\mathrm{ov}}$, 
but now $\Delta F_{\mathrm{az}}$ is removing the freshwater anomalies from the latitude band.
The surface flux $F_{\mathrm{surf}}$ is slightly declining prior to the strong increase in $\Delta F_{\mathrm{ov}}$,
which implies that the salt-advection feedback is triggered by heat  fluxes (Figure~\ref{fig:Figure_S2}g).
The freshwater accumulation over the upper 1,000~m ($\overline{W}_{1000\uparrow}$) is dominant in the faster CO$_2$ ramps and 
has a different sign when comparing to the slow CO$_2$ ramp (Figures~\ref{fig:Figure_4}g and \ref{fig:Figure_S4}). 
This dichotomy is also evident among the different RCP scenarios (Figure~\ref{fig:Figure_4}h). 
Lower values of $\overline{W}_{1000\uparrow}$ are therefore indicative of increased AMOC stability \cite{Dai2025},
 but should be considered with care as they can be compensated by temperature responses (Figure~\ref{fig:Figure_S4}a).

As the AMOC remains stable in the slow CO$_2$ ramp simulation, it is also useful to analyse the entire Atlantic freshwater budget between 34$^{\circ}$S and 65$^{\circ}$N.
The integrated freshwater budget can provide insights in the dominant feedbacks \cite{Huisman2010, Vanderborght2025}, such as the strength of the salt-advection feedback which is
quantified by the AMOC-induced freshwater transport at 34$^{\circ}$S, i.e., $F_{\mathrm{ovS}}$.  
Using $F_{\mathrm{ovS}}$ as a stability indicator is only appropriate under (quasi-)equilibrium conditions \cite{Rahmstorf1996,Vanderborght2025}, 
which certainly does not apply for the faster CO$_2$ ramps (inset in Figure~\ref{fig:Figure_1}d) nor the RCP4.5 and RCP8.5 scenarios \cite{vanWesten2025e}.
The global warming rate in the slow CO$_2$ ramp simulation is on average 0.03$^{\circ}$C per decade, 
about a factor of 10~slower compared to that in the CO$_2$ ramp of +5.0~ppm~yr$^{-1}$, the RCP8.5 scenario over the 21st century, and the currently observed warming rate \cite{Hansen2023,Foster2026}.

\begin{figure}[h]

\begin{tabular}{c}

\includegraphics[width=1\columnwidth, trim = {0cm 0cm 0cm 0cm}, clip]{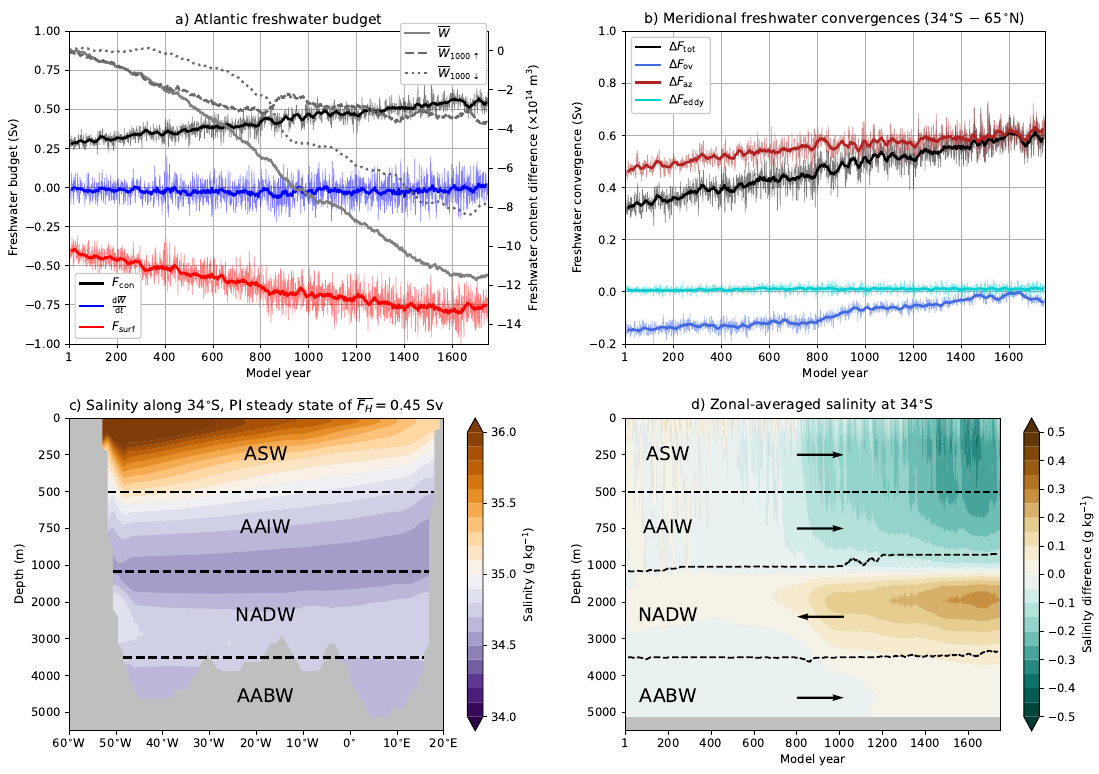} 
\end{tabular}

\caption{\textbf{The Atlantic freshwater budget} 
(a): The Atlantic Ocean (34$^{\circ}$S to 65$^{\circ}$N) freshwater budget for the slow CO$_2$ ramp
with the freshwater content ($\overline{W}$), freshwater convergence ($F_{\mathrm{con}}$),
surface freshwater fluxes ($F_{\mathrm{surf}}$) and changes in the freshwater content ($\frac{\mathrm{d}\overline{W}}{\mathrm{d}t}$).
The quantity $\overline{W}$ is split into an upper 1,000~m contribution ($\overline{W}_{1000\uparrow}$) and below 1,000~m contribution ($\overline{W}_{1000\downarrow}$)
and are displayed as their differences compared to PI$_{45}^{\mathrm{on}}$.
(b): The meridional freshwater convergences (omitting the contribution by the Strait of Gibraltar) for the different freshwater transport components.
(c): The salinity along 34$^{\circ}$S for PI$_{45}^{\mathrm{on}}$, 
where the dashed lines indicate the different water masses which are based on the meridional velocity profile (see procedure outlined in \cite{vanWesten2024b}).
(d): The zonally-averaged salinity at 34$^{\circ}$S over time, which are displayed as differences compared to PI$_{45}^{\mathrm{on}}$.
The dashed lines are the different water masses.
The arrows are indicative of the meridional velocity (right = northward, left = southward) associated with the AMOC.
}

\label{fig:Figure_5}

\end{figure}

The Atlantic Ocean is a net evaporative basin. In the slow CO$_2$ ramp simulation, this characteristic becomes more pronounced in warmer climates (Figure~\ref{fig:Figure_5}a). While precipitation increases over most of the ocean, indicative of an enhanced hydrological cycle, evaporation likewise increases over all ocean surfaces (Figures~\ref{fig:Figure_S5}c,e). Over the Atlantic basin, this leads to a net negative precipitation minus evaporation ($P-E$) anomaly (i.e., salinifying, Figures~\ref{fig:Figure_S5}a--e), contrary to the conventional thought that global warming increases $P-E$ over the North Atlantic.
A portion of these freshwater anomalies is transferred via atmospheric bridges to other basins.
Note that there is also a considerable sea-ice contribution at the higher latitudes (Figures~\ref{fig:Figure_S5}f).
Hence, surface freshwater fluxes drive the salinifying Atlantic Ocean (lower $\overline{W}$), 
which is partially balanced by horizontal freshwater convergences (Figure~\ref{fig:Figure_5}a). 

The freshwater convergence (over 34$^{\circ}$S -- 65$^{\circ}$N) is again decomposed into its dominant contributing factors (Figure~\ref{fig:Figure_5}b). 
While $\Delta F_{\mathrm{az}}$ is mostly contributing before model year~800, $\Delta F_{\mathrm{ov}}$ thereafter also contributes.
 The increase in $\Delta F_{\mathrm{az}}$ is primarily caused by a declining gyre freshwater transport at 65$^{\circ}$N over the first 800~model years, after which  $\Delta F_{\mathrm{az}}$ stabilises (Figure~\ref{fig:Figure_S6}b).
This response appears to be related to the retreating North Atlantic sea-ice extent. 
The freshwater transport by the overturning component at 34$^{\circ}$S, $F_\mathrm{ovS}$, starts to contribute after model year~800 (Figure~\ref{fig:Figure_S6}a),
which is also reflected by the salinity responses at 34$^{\circ}$S (Figures~\ref{fig:Figure_5}c,d).
The southward flowing salinity anomalies between 1000 -- 3500~m depths, which are part of the North Atlantic Deep Water (NADW), have a North Atlantic origin (Figure~\ref{fig:Figure_4}b).
The latter is crucial, as it indicates changes in the inter-hemispheric meridional salinity contrast, 
which can be linked to AMOC strength and stability through $F_{\mathrm{ovS}}$ \cite{Vanderborght2025}.
The $F_{\mathrm{ovS}}$ is increasing (Figure~\ref{fig:Figure_S6}c), meaning that the salt-advection feedback becomes weaker. 
An increased $F_{\mathrm{ovS}}$ variance is also indicative of a stronger salt-advection feedback and hence a less 
stable AMOC \cite{vanWesten2024a,vanWesten2025b}. In the slow CO$_2$ ramp simulation, the  $F_{\mathrm{ovS}}$ variance declines (Figure~\ref{fig:Figure_S6}d), indicating AMOC stability changes on multi-centennial timescales. 
For shorter timescales, however, such as during rapid climate change, it is more useful to analyse AMOC-related quantities over the North Atlantic Ocean (40$^{\circ}$N -- 65$^{\circ}$N; Figures~\ref{fig:Figure_3} and \ref{fig:Figure_4}), 
as the inter-hemispheric meridional salinity contrast has insufficient time to adjust.

The salinifying NADW is an important fingerprint of increased AMOC stability and is only relevant in cases with a sufficiently strong AMOC, 
allowing the inter-hemispheric salinity contrast to adjust.
This fingerprint is found under the RCP4.5 scenario with $\overline{F_H} = 0.18$~Sv background hosing, 
but not for RCP2.6 and $\overline{F_H} = 0.45$~Sv (Figure~\ref{fig:Figure_S7}), even though the AMOC avoids a collapse in both cases. 
This difference could be related to the different $\overline{F_H}$ values; however, it is more likely related to the GMST anomaly. 
Higher temperature anomalies induce more evaporation which eventually results in a saltier NADW; this is indeed the case for the slow CO$_2$ ramp (Figure~\ref{fig:Figure_5}d).
The NADW salinity anomalies become more pronounced after model year~700, corresponding to GMST anomalies above +2.5$^{\circ}$C in the slow CO$_2$ ramp. 
The simulation under RCP4.5 and $\overline{F_H} = 0.18$~Sv exceeds this warming level around model year~2200 and thereafter NADW salinities increase, 
whereas the RCP2.6 and $\overline{F_H} = 0.45$~Sv simulation stays below +2$^{\circ}$C warming. 

We also test whether a salinification of NADW occurs in other models under emission scenarios with a sustained AMOC. 
Indeed, the fingerprint is visible in the GISS-E2-1-G model under the extended SSP1-2.6 and SSP2-4.5 scenarios (Figure~\ref{fig:Figure_S8}), showing larger salinity anomalies under SSP2-4.5 than SSP1-2.6.
Under the extended SSP1-2.6 scenario, most other CMIP6 models show an increase in NADW salinity between 1000 -- 2000~m depths (Figure~\ref{fig:Figure_S9}),
but these responses are quite small and the NADW even becomes fresher in the IPSL-CM6A-LR model.
While we expect a stronger NADW fingerprint at higher radiative forcing, this cannot be verified as all CMIP6 models show a collapsing AMOC under the high-emission SSP5-8.5 scenario \cite{Drijfhout2025}.
Although the signal-to-noise ratio is relatively small, most CMIP6 models align qualitatively with our CESM results, suggesting increased AMOC stability in warmer climates.
The near-surface freshening at 34$^{\circ}$S (Figure~\ref{fig:Figure_5}d) is also contributing to increasing $F_{\mathrm{ovS}}$ values in the slow CO$_2$ ramp (Figure~\ref{fig:Figure_S6}c), but there is a large inter-model spread, likely related to differences in ocean-atmosphere interactions, Aghulas Leakage, and model resolution used \cite{vanWesten2024b, Grosselindemann2025}.
Consequently, the destabilising effects of anthropogenic climate change at high enough forcing rate and can become 
larger than the stabilising responses, such that the AMOC is brought out of equilibrium and subsequently collapses.

Overall, the freshwater budget analysis in the CESM and CMIP6 models supports the idea that the present-day AMOC state becomes more stable in a warmer climate. This suggests that there is no specific GMST threshold for AMOC tipping (in the examined global warming range). To cause an AMOC collapse, the rate-dependent destabilising effects of anthropogenic forcing must overcome the slow stabilising effects. Consequently, we conclude that the AMOC collapse observed in the faster CO$_2$ ramps and higher RCP scenarios is rate-induced: it occurrs due to exceeding a critical warming rate, without crossing a critical threshold in global warming level.

\section*{Dynamical Perspective}

It remains unclear whether the AMOC could cross a saddle-node bifurcation under anthropogenic climate change. Addressing this question 
requires additional quasi-equilibrium simulations in the forcing space of imposed freshwater fluxes and radiative forcing.
Moreover, the AMOC response depends on the background climate state, e.g. on $\overline{F_H}$ \cite{vanWesten2025e}, and
sensitivity experiments with models like the CESM require substantial computational resources.
Therefore, we here use a 5-box ocean model of the Atlantic basin (Figure~\ref{fig:Figure_S10}a) to conceptualize the forcing path dependence of AMOC tipping.

The box model has been shown to capture the essential AMOC dynamics of the CESM \cite{vanWesten2025a}. In our context, the two main forcing parameters are the freshwater flux forcing ($E_A$) and the atmospheric temperature anomaly over the subpolar Atlantic box ($\Delta T_{\mathrm{n}}^a$).
The effects of global warming are incorporated by increasing $\Delta T_{\mathrm{n}}^a$ to lower the meridional atmospheric temperature gradient between the lower and higher latitudes, 
mimicking polar amplification \cite{vanWesten2025a}. 
For each $\Delta T_{\mathrm{n}}^a$ increment of 0.1$^{\circ}$C, we calculate the steady states and bifurcations of the box model 
using continuation techniques, with $E_A$ as the bifurcation parameter.
The model features a bistable AMOC regime bounded by saddle-node bifurcations, which can be crossed in both forcing directions ($E_A$ and $\Delta T_{\mathrm{n}}^a$).
The saddle-node bifurcation shifts in freshwater forcing space from $E_A = 0.486$~Sv for $\Delta T_{\mathrm{n}}^a = 0^{\circ}$C 
to  $E_A = 0.342$~Sv for $\Delta T_{\mathrm{n}}^a = +5^{\circ}$C (Figures~\ref{fig:Figure_S10}b).

\begin{figure}[h!]

\begin{tabular}{c}

\hspace{-2cm}
\includegraphics[width=1.3\columnwidth, trim = {0cm 0cm 0cm 0cm}, clip]{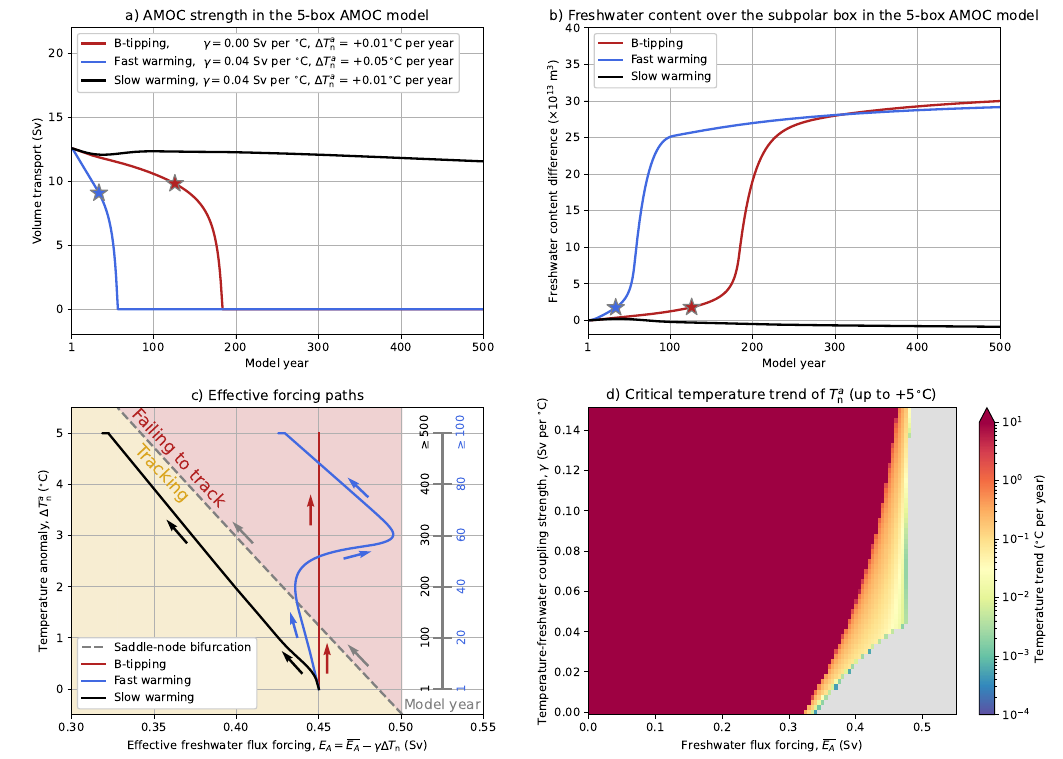} 
\end{tabular}

\caption{\textbf{AMOC transitions in the 5-box model} 
(a \& b): The AMOC strength and the freshwater content difference (compared to initial state) over the subpolar box for three different forcing paths (see legend in panel~a for details),
all forcing paths have $\overline{E_A} = 0.45$~Sv.
The stars indicate the crossing of the saddle-node bifurcation.
(c): The position of the saddle-node bifurcation of the stable AMOC on state for increasing $\Delta T_{\mathrm{n}}^a$, including the three forcing paths.
(d): The critical temperature trend of $T_{\mathrm{n}}^a$ without collapsing the AMOC for varying $\overline{E_A}$ and temperature-freshwater coupling strength $\gamma$,
the lowest explored trend is $1 \times 10^{-4}~^{\circ}$C per year.} 

\label{fig:Figure_6}

\end{figure}

First, consider cases with fixed $E_A > 0.342$~Sv. The AMOC always collapses under an imposed warming anomaly of $\Delta T_{\mathrm{n}}^a = +5^{\circ}$C or more, since the `AMOC on' state disappears. Thus, a critical warming threshold exists beyond which the system transitions to the only remaining stable solution (the collapsed `AMOC off' state), regardless of the rate of increase at which the warming anomaly is reached. This is a case of bifurcation-induced tipping (B-tipping).
An example is given for a fixed freshwater flux forcing ($\overline{E_A}$) of 0.45~Sv, forced under a linear temperature trend of $\Delta T_{\mathrm{n}}^a = +0.01^{\circ}$C per year up to +5$^{\circ}$C (red curve in Figure~\ref{fig:Figure_6}a).

This situation is qualitatively inconsistent with the slow CO$_2$ ramp simulation in CESM, where the `AMOC on' state remains stable up to at least $+5.5^\circ$C. However, the box model so far neglects the inter-dependence between surface heat and freshwater fluxes under radiative forcing change. To account for this in an idealized manner, we consider an effective freshwater flux forcing $E_A$ that depends on the ocean state, specifically the temperature anomaly $\Delta T_{\mathrm{n}}$ of the subpolar box: $E_A = \overline{E_A} - \gamma \Delta T_{\mathrm{n}}$.
Here $\overline{E_A}$ is the fixed freshwater flux forcing, $\gamma$ is the temperature--freshwater coupling strength, and $\Delta T_{\mathrm{n}} = T_{\mathrm{n}}(t) - T_{\mathrm{n}}(\overline{E_A},\Delta T_{\mathrm{n}}^a = 0)$.
This temperature--freshwater coupling is motivated by the declining North Atlantic surface freshwater input in the slow CO$_2$ ramp CESM simulation (Figures~\ref{fig:Figure_4}a and \ref{fig:Figure_5}a)
where, for reference, the freshwater flux over 34$^{\circ}$S -- 65$^{\circ}$N declines by 0.069~Sv per degree warming (giving $\gamma \approx 0.069$~Sv per $^{\circ}$C).

We repeat the warming experiment (linear ramp up to $\Delta T_{\mathrm{n}}^a = 5^\circ$C at fixed $\overline{E_A} = 0.45$~Sv) but now with $\gamma > 0$, comparing relatively slow (+0.01$^{\circ}$C yr$^{-1}$) and relatively fast (+0.05$^{\circ}$C yr$^{-1}$) warming rates.
For $\gamma = 0.04$~Sv per $^{\circ}$C, the system continues to track the `AMOC on' state under the slow warming case  (black curve in Figure~\ref{fig:Figure_6}a).
The equilibrated subpolar ocean warming $\Delta T_{\mathrm{n}}$ is 3.17$^{\circ}$C, resulting in an effective hosing strength of 0.32~Sv,
which is indeed below the bifurcation threshold of 0.342~Sv for $\Delta T_{\mathrm{n}}^a = +5^{\circ}$C.
However, in the fast warming case, the AMOC collapses as the system fails to track the `AMOC on' state after crossing the moving saddle-node bifurcation (blue curve in Figure~\ref{fig:Figure_6}a).
Since the `AMOC on' state remains stable for $\gamma = 0.04$~Sv per $^{\circ}$C (Figures~\ref{fig:Figure_S10}c,d) and can be tracked in the slow warming case, the fast warming case represents a rate-induced AMOC tipping event.

To visualise the cases described above, we plot their effective forcing paths in a diagram of the warming anomaly against the effective freshwater flux forcing (Figure~\ref{fig:Figure_6}c).
For $\gamma > 0$, trajectories initially move towards lower effective freshwater flux forcing, and whether the saddle-node bifurcation is crossed depends on the forcing rate.
Tracking the stable `AMOC on' state hence depends on the ratios of atmospheric warming rate (forcing timescale) and ocean warming rate (adjustment timescale), 
as well as the interaction strength between thermal and freshwater effects.
Due to the state-dependent surface forcing owing to ocean-atmosphere coupling, the rate-dependent tipping scenario shown here is more complex than the classical setting of rate-induced tipping 
in which the control parameters are purely external \cite{Ashwin2012,Wieczorek2023}.

The box model is a strong simplification of the CESM dynamics and only mimicks one of the rate-dependent mechanisms identified in the CESM, namely the reduced freshwater flux through the Atlantic Ocean surface under gradual warming. Nonetheless, this process is sufficient to produce rate-dependent AMOC tipping in the box model, and the freshwater content in the subpolar box shows a qualitatively similar response to the freshwater content analysis in the CESM (Figures \ref{fig:Figure_6}b and \ref{fig:Figure_4}g). The maximal temperature trend that does not cause an AMOC collapse (i.e., the critical warming rate), is shown in Figure~\ref{fig:Figure_6}d as a function of $\overline{E_A}$ and $\gamma$. 
For sufficiently large temperature-freshwater coupling strengths, the system can always track the stable `AMOC on' state as long as it exists and the imposed forcing rate is slow enough. 
These results are qualitatively robust for even larger temperature anomalies of up to +10$^{\circ}$C (Figure~\ref{fig:Figure_S10}e), 
demonstrating that even for extreme warming, forcing pathways may exist for which the AMOC does not collapse.

\section*{Discussion}

In this study, we argue that an AMOC collapse is dependent on the radiative forcing path (i.e., climate change scenario) and not governed by a specific global warming threshold. 
To demonstrate this, we presented an analysis of CO$_2$ ramp simulations using the Community Earth System Model, where the CO$_2$ concentration was increased up to four times preindustrial levels at different rates.
In the slow CO$_2$ ramp (+0.5~ppm~yr$^{-1}$) simulation, the GMST anomaly reaches +5.5$^{\circ}$C and the AMOC remains stable throughout,  
while for the faster CO$_2$ ramps (+2.5~ppm~yr$^{-1}$ and +5.0~ppm~yr$^{-1}$) the AMOC starts to collapse at +2.0$^{\circ}$C.
This GMST anomaly is much lower than the maximum warming in the slow CO$_2$ ramp, showing that an AMOC collapse is determined by the radiative forcing path and specifically the forcing rate. 
Comparable results are found under the standard climate change scenarios of RCP2.6 (stable AMOC) and RCP4.5 and RCP8.5 (collapsing AMOC),
where we note that these RCPs have different forcing agent (CO$_2$, methane, aerosols) pathways that all affect the AMOC \cite{Menary2020,Hassan2021,Hankel2025}.
Our result corroborates early work using idealised climate models \cite{Stocker1997}, while now being demonstrated with a much more comprehensive climate model. 

Present-day and projected warming simulations using CMIP6 models show substantial AMOC weakening \cite{Caesar2018, 
Weijer2020, Bonan2025, Michel2025, Dijkstra2026} and possibly an AMOC collapse \cite{Romanou2023,Drijfhout2025}.
For the CESM under relatively rapid radiative forcing changes ($\geq$ +2.5~ppm~yr$^{-1}$ and $\overline{F_H} = 0.45$~Sv), 
the near-surface water is getting lighter much faster than the interior of the Atlantic Ocean, 
resulting in a shutdown of the adiabatic pathways followed by an AMOC collapse.
When the imposed radiative forcing is sufficiently slow, 
both the interior and surface waters adjust their densities accordingly such that they support a relatively strong adiabatic AMOC. 
With an idealised AMOC model subjected to heat and freshwater forcing, 
we illustrated how the interplay between the changing forcing and the changing climate state can alter the system's stability landscape in a way that leads to rate-dependent AMOC tipping.

In the slow CO$_2$ ramp simulation, the relatively small temperature-induced AMOC weakening is outweighed by two dominant effects. 
First, higher evaporation rates make the evaporative characteristic of the Atlantic basin more pronounced, which effectively results in a negative freshwater flux forcing.
Second, the North Atlantic sea-ice extent is reduced in a warmer climate, which limits the sea-ice insulation effects that 
influence the AMOC \cite{Lin2023, vanWesten2024c} and also lowers the freshwater input by sea-ice melt, contributing to less freshwater input over the higher latitudes.
Both net evaporation and reduced sea-ice extent contribute to a salinifying Atlantic Ocean 
and, after water mass transformation, the AMOC starts to export these salinity anomalies out of the Atlantic basin at greater depth (1000 -- 3500~m). 
This response at depth is indicative of a more stable AMOC (i.e., larger  $F_{\mathrm{ovS}}$ values) 
and a similar response is found in most CMIP6 models under extended SSP scenarios with a sustained AMOC. 
The freshwater budget and the related $F_{\mathrm{ovS}}$ stability indicator are only relevant under 
(quasi-)equilibrium conditions with a stable AMOC state \cite{Rahmstorf1996, Vanderborght2025} and are not useful 
under rapid climate change \cite{vanWesten2025e}.
These stabilising mechanisms explain the strong AMOC states found in warmer climates \cite{Bonan2022,Romanou2023,Willeit2024b,Hankel2025}. 
Some models, such as the CESM version used here, are in a monostable regime under PI conditions without background hosing \cite{vanWesten2023b} and
this is related to persistent climate model biases \cite{vanWesten2024b,Dijkstra2024b,Boot2025}, 
which could also be the case under different radiative forcing conditions.

There are now first indications that the AMOC can collapse under anthropogenic climate change in the latest generation 
of climate models \cite{Romanou2023,Drijfhout2025}. This implies that the stabilising mechanism in warmer 
climates, as described above, operates on timescales slower than the current rate of radiative forcing. 
Hence, it may still be useful to explore warming levels for an AMOC collapse onset on relatively 
short timescales (years to decades), but this becomes less useful if warming rates slow down (as projected beyond the 21st century) and should be complemented with threshold estimates for other relevant forcing parameters.
Based on our results, we propose to convert the AMOC tipping warming threshold to a warming rate threshold. 
Yet, this also poses challenges as the critical warming rate is dependent on the background climate state and the strength of the stabilising mechanism.
Therefore, it is more useful to analyse physics-based indicators, such as $\Psi_{\mathrm{NADW}}$ and $B_{\mathrm{flux}}$,
that capture the effects of changing forcing conditions (both stabilising and destabilising).
The distance to the tipping point could be estimated using $B_{\mathrm{flux}}$, 
with a $B_{\mathrm{flux}}$ sign change marking the onset of an AMOC collapse also under transient climate change \cite{vanWesten2025e}.

According to the $B_{\mathrm{flux}}$ indicator, as determined in CMIP6 models, the AMOC starts to collapse around +2.5$^{\circ}$C warming at projected warming rates, which could be reached around the year 2060 \cite{vanWesten2025e}. 
This is substantially lower than the previous estimate of +4$^{\circ}$C warming \cite{Armstrong2022} that lacks a time horizon. 
Observations show a present-day warming close to the +1.5$^{\circ}$C level (2023 -- 2025, C3S/ECMWF and \cite{Foster2026}) and, 
if the AMOC were to begin collapsing at +2.5$^{\circ}$C warming in 2060, this would imply a critical warming rate of +0.29$^{\circ}$C per decade.
This rate aligns with projected warming rates of +0.27 to +0.36$^{\circ}$C per decade \cite{Hansen2023} and,
also considering the effects of random variability \cite{Slyman2023}, strongly suggests that rate-dependent effects are highly relevant for the fate of the AMOC.
It is therefore critically important to reduce the rate of radiative forcing as quickly as possible to limit the risk of  an AMOC collapse.

\newpage  
\backmatter

\section*{Methods}

\bmhead{Climate Model Simulations} 

The CESM is a fully-coupled climate model and the simulations here have a 1$^{\circ}$ horizontal resolution for the ocean/sea-ice components and a 2$^{\circ}$ horizontal resolution for the atmosphere/land components.  
The ocean component is the Parallel Ocean Program version~2 (POP2, \cite{Smith2010}),
the atmospheric component is the Community Atmosphere Model version~4 (CAM4, \cite{Neale2013}), 
and the sea-ice component is the Community Ice Code version~4 (CICE4, \cite{Hunke2008}). 
The CESM is either forced under varying freshwater flux forcing and fixed pre-industrial radiative forcing conditions 
or fixed freshwater flux forcing and varying radiative forcing conditions.
For more details on the precise CESM set-up, we refer to previous work \cite{vanWesten2023b,vanWesten2024a,vanWesten2024c,vanWesten2025e} and the main text.\\

In addition to the CESM, model output from CMIP phase~6 (indicated here as CMIP6 models) were analysed. 
We retained the historical forcing (1850 -- 2014) followed by the extended SSP1-2.6 scenario (2015 -- 2300). 
Only for the GISS-E2-1-G model, the scenarios were extended to 2500 and are also available for the extended SSP2-4.5 scenario.
Note that the forcing scenarios are slightly different between the CESM simulations (the RCPs) and CMIP6 (the SSPs).

\bmhead{Water Mass Transformation (WMT)}

We use the same procedure as described in \cite{vanWesten2025e}, which is briefly repeated here for completeness.
We consider the volume ($V_\sigma$) conservation of a fixed horizontal domain and that is bounded by the ocean surface and isopycnal \cite{Groeskamp2019}:
\begin{equation}
\frac{\partial V_\sigma}{\partial t} = M_\sigma - G_\sigma ,
\end{equation}
with $M_\sigma$ is the advective transport convergence through 
the open boundaries of the domain and $G_\sigma$ is the diapycnal 
transformation rate taken place in the domain.
The WMT rates from the surface are dominant in $G_\sigma$ and are indicated by $\Psi_{\mathrm{WMT}}$:
\begin{equation}
\Psi_{\mathrm{WMT}}(y,\sigma_2) = \frac{1}{\Delta  \sigma_2} \int_{x_W}^{x_E} \int_y^{y_N} -\frac{\rho_0}{g} B_{\mathrm{flux}}(x,y)~\Pi(\sigma_2)~\mathrm{d} y' \mathrm{d} x'
\end{equation}
where:
\begin{equation}
 \Pi(\sigma_2) =
\begin{cases}
1~\hspace{0.5cm}~\text{if}~\sigma_2 - \frac{\Delta \sigma_2}{2} \leq \sigma_2 <  \sigma_2 + \frac{\Delta \sigma_2}{2} \\
0~\hspace{0.5cm}~\text{elsewhere}
\end{cases}
\end{equation}
and we use a potential density bin size of $\Delta \sigma_2 = 0.05$~kg~m$^{-3}$.
The quantity $B_{\mathrm{flux}}$ is the surface buoyancy flux:
\begin{equation} \label{eq:B_flux}
B_{\mathrm{flux}}(x,y) = \frac{g \alpha}{\rho_0 C_p} Q_{\mathrm{heat}} + \frac{g \beta S_{\mathrm{surf}} }{\rho_0} Q_{\mathrm{fresh}} = B_{\mathrm{flux}}^T + B_{\mathrm{flux}}^S
\end{equation}
and we also analyse the upper 1,000~m ocean buoyancy:
\begin{equation}
B_{\mathrm{ocean}}(T, S) = \frac{g}{\rho_0} \int_{-1000}^{0} \left( \sigma(z) - \sigma(z=0) \right) \mathrm{d}z
\end{equation}
In the relations above, $\alpha$ is the thermal expansion coefficient, $C_p$ the specific heat capacity, $Q_{\mathrm{heat}}$ the net heat flux into the ocean,
$\beta$ the haline contraction coefficient, $S_{\mathrm{surf}}$ the sea surface salinity, $Q_{\mathrm{fresh}}$ the net freshwater flux into the ocean,
$g$ (= 9.8~m~s$^{-2}$) is the gravitational acceleration and $\rho_0$ (= 1027~kg~m$^{-3}$) a reference density. 
For each quantity, we used its local value and for $\alpha$, $\beta$ and $C_p$ we used the Thermodynamic Equation of SeaWater 2010 (TEOS-10) toolkit \cite{Mcdougall2011}. 
All the analyses are conducted on monthly-averaged fields (due to strong seasonal cycle) and the related time series are subsequently averaged to yearly values.

\bmhead{Defining the onset of AMOC collapse} 

An AMOC collapse threshold can be accurately determined for the quasi-equilibrium PI hosing simulation, where the crossing of the tipping point coincides in time with the start of abrupt AMOC weakening \cite{vanWesten2024a}.
However, under transient climate change, it is more challenging to deduce the onset of the collapse, since the transition is not sharp in time and is preceded by AMOC weakening \cite{Drijfhout2025}.
\citet{vanWesten2025e} proposed the sign change in the spatially-averaged $B_{\mathrm{flux}}$ over the North Atlantic isopycnal outcropping region (40$^{\circ}$N -- 65$^{\circ}$N) as an estimate of the AMOC collapse onset (Figures~\ref{fig:Figure_S2}g,h),
with the timing of the sign change being robust when this latitudinal band is slightly varied.
The spatially-averaged $B_{\mathrm{flux}}$ is strongly related to the adiabatic pathways that are part of the AMOC (i.e., $\Psi_{\mathrm{NADW}}$, see main text),
and this estimate works very well for CESM, as well as for other CMIP6 models \cite{vanWesten2025e}.

Here, the AMOC collapse onset is defined as the time point (or corresponding GMST anomaly) at which the 11-year running mean of $B_{\mathrm{flux}}$ first switches sign from negative to positive (Table~\ref{tab:Table_S1}). This definition holds whether the tipping event is rate-induced or bifurcation-induced.

\bmhead{The Atlantic Freshwater Budget}

The freshwater budget over a latitude band ($y_1$ to $y_2$) in the Atlantic Ocean is defined as:

\begin{subequations}
\begin{align}
\frac{\mathrm{d}\overline{W}}{\mathrm{d}t} &= F_\mathrm{con} + F_\mathrm{surf} + F_\mathrm{mix} \\
\overline{W}  &=  - \frac{1}{S_0} \int_{-H}^0 \int_{y_1}^{y_2} \int_{x_W}^{x_E} (S - S_0) \mathrm{d} x \mathrm{d} y \mathrm{d}z
\end{align}
\end{subequations}
where $\overline{W}$ is the freshwater content, $F_\mathrm{con}$ the freshwater convergence and is determined as the total freshwater transport through the boundaries,
$F_\mathrm{surf}$ the surface freshwater flux and $F_\mathrm{mix}$ is a residual term which closes the budget and captures for example diffusion \cite{Juling2021}. 
The $F_\mathrm{con}$ is primarily governed by the total meridional freshwater transports and is defined by:
\begin{equation}
F_\mathrm{tot}(y) = - \frac{1}{S_0} \int_{-H}^0 \int_{x_W}^{x_E} v (S - S_0) \mathrm{d} x \mathrm{d}z 
\end{equation}
and can be further decomposed in an overturning ($F_\mathrm{ov}$), azonal gyre ($F_\mathrm{az}$), barotropic ($F_\mathrm{bt} \approx 0$), and eddy ($F_\mathrm{eddy}$) contribution:
\begin{eqnarray}
F_\mathrm{ov}(y)   &= & - \frac{1}{S_0} \int_{-H}^0 \left[ \int_{x_W}^{x_E} v^* \mathrm{d} x \right] \left[ \langle S \rangle - S_0 \right] \mathrm{d}z ,\\
F_\mathrm{az}(y)  &= & - \frac{1}{S_0} \int_{-H}^0 \int_{x_W}^{x_E} v' S' \mathrm{d} x \mathrm{d}z ,\\
F_\mathrm{bt}(y)  &= & - \frac{1}{S_0} \int_{-H}^0 \int_{x_W}^{x_E} \hat{v} \left( \hat{S} - S_0 \right) \mathrm{d} x \mathrm{d}z ,\\
F_\mathrm{eddy}(y)  &= & - \frac{1}{S_0} \int_{-H}^0 \int_{x_W}^{x_E} \tilde{v} \left( \tilde{S} - S_0 \right) \mathrm{d} x \mathrm{d}z .
\end{eqnarray}
Here, $v^*$ is defined as $v^* = v - \hat{v}$,
where $v$ is the meridional velocity and $\hat{v}$ ($\hat{S}$) is the section spatially-averaged meridional velocity (salinity).
The quantity $\langle S \rangle$ indicates the zonally-averaged salinity and primed quantities ($v'$ and $S'$) are deviations from their respective zonal averages.
The $\tilde{v}$ and $\tilde{S}$ are the eddy terms for velocity and salinity, respectively, with more details provided in \cite{Juling2021}.

\bmhead{Acknowledgments}

The model simulation and the analysis of all the model output was conducted on the Dutch National 
Supercomputer Snellius within NWO-SURF project 2024.013. 
The authors thank Michael Kliphuis (IMAU, UU) for performing these simulations,
and Paul Ritchie (University of Exeter) for the useful discussion on rate-induced tipping.

\section*{Declarations}

\begin{itemize}
\item Funding -- R.M.v.W. and H.A.D. are funded by the European Research Council through the ERC-AdG project TAOC (project 101055096). R.B. greatfully acknowledges
the ClimTip project, which has received funding from the European Union's Horizon Europe research and innovation programme under grant agreement No. 101137601. This is ClimTip contribution \#150.
\item Conflict of interest -- The authors declare no competing interest
\item Ethics approval -- Not applicable
\item Availability of data and materials -- The (processed) model output are available at:  https://doi.org/10.5281/zenodo.19853635 . 
The CMIP6 model output is provided by the World Climate Research Programme’s Working Group on Coupled Modeling. 
\item Code availability -- The analysis scripts are also available at: https://doi.org/10.5281/zenodo.19853635
\item Authors' contributions -- R.M.v.W., R.B.  and H.A.D. conceived the idea for this study.
R.M.v.W. conducted the analysis and prepared all figures. All  authors were actively involved in the interpretation of the analysis results and the writing process.
\end{itemize}

\newpage 
\setcounter{figure}{0}

\newpage 
\setcounter{figure}{0}

\makeatletter 
\renewcommand{\thefigure}{S\@arabic\c@figure}
\makeatother

\makeatletter 
\renewcommand{\thetable}{S\@arabic\c@table}
\makeatother


\section*{Supplementary Table}

\begin{table}[h]
\caption{Overview of the available CESM simulations, including the value of $\overline{F_H}$ and forcing conditions.
Where applicable, the timing of AMOC collapse onset and the associated AMOC strength and GMST anomaly (compared to: 1 = Historical of $\overline{F_H} = 0.18$~Sv (1850 -- 1899); 2 = Historical of $\overline{F_H} = 0.45$~Sv (1850 -- 1899); 3 = PI steady state of $\overline{F_H} = 0.45$~Sv)
are also indicated.}

\hspace{-2cm}
\begin{tabular}{|c|c|c|c|c|c|c|c|}
\hline
Simulation	 name						& $\overline{F_H}$			& Forcing conditions				& Tipping time	& AMOC strength	& GMST anomaly	\\
									& (Sv)					& 							& (model year)	& (Sv)			& ($^{\circ}$C) 		\\ \hline \hline
Extended RCP4.5 \#1					& \multirow{2}{*}{0.18}		& Historical--RCP4.5				& --			& --				& -- 				 \\
Extended RCP8.5 \#1					& 						& Historical--RCP8.5				& 2099		& 7.8				& $+4.15^1$		 \\ \hline \hline
Extended RCP2.6						& \multirow{3}{*}{0.45}		& Historical--RCP2.6				& --			& --				& --				 \\
Extended RCP4.5 \#2					& 						& Historical--RCP4.5				& 2111		& 8.3				& $+2.19^2$		 \\
Extended RCP8.5 \#2					& 						& Historical--RCP8.5				& 2061		& 9.0				& $+2.78^2$		 \\ \hline \hline
CO$_2$ ramp \#1						& \multirow{3}{*}{0.45}		& +0.5 ppm yr$^{-1}$ of CO$_2$	& --			& --				& --				 \\
CO$_2$ ramp \#2						& 						& +2.5 ppm yr$^{-1}$ of CO$_2$	& 139		& 9.0				& $+2.02^3$		 \\
CO$_2$ ramp \#3						& 						& +5.0 ppm yr$^{-1}$ of CO$_2$	& 78			& 8.8				& $+1.99^3$		 \\ \hline

\end{tabular}
\label{tab:Table_S1}
\end{table}

\section*{Supplementary Figures}


\begin{figure}[h]

\hspace{-3cm}

\begin{tabular}{c}

\includegraphics[width=1\columnwidth, trim = {0cm 0cm 0cm 0cm}, clip]{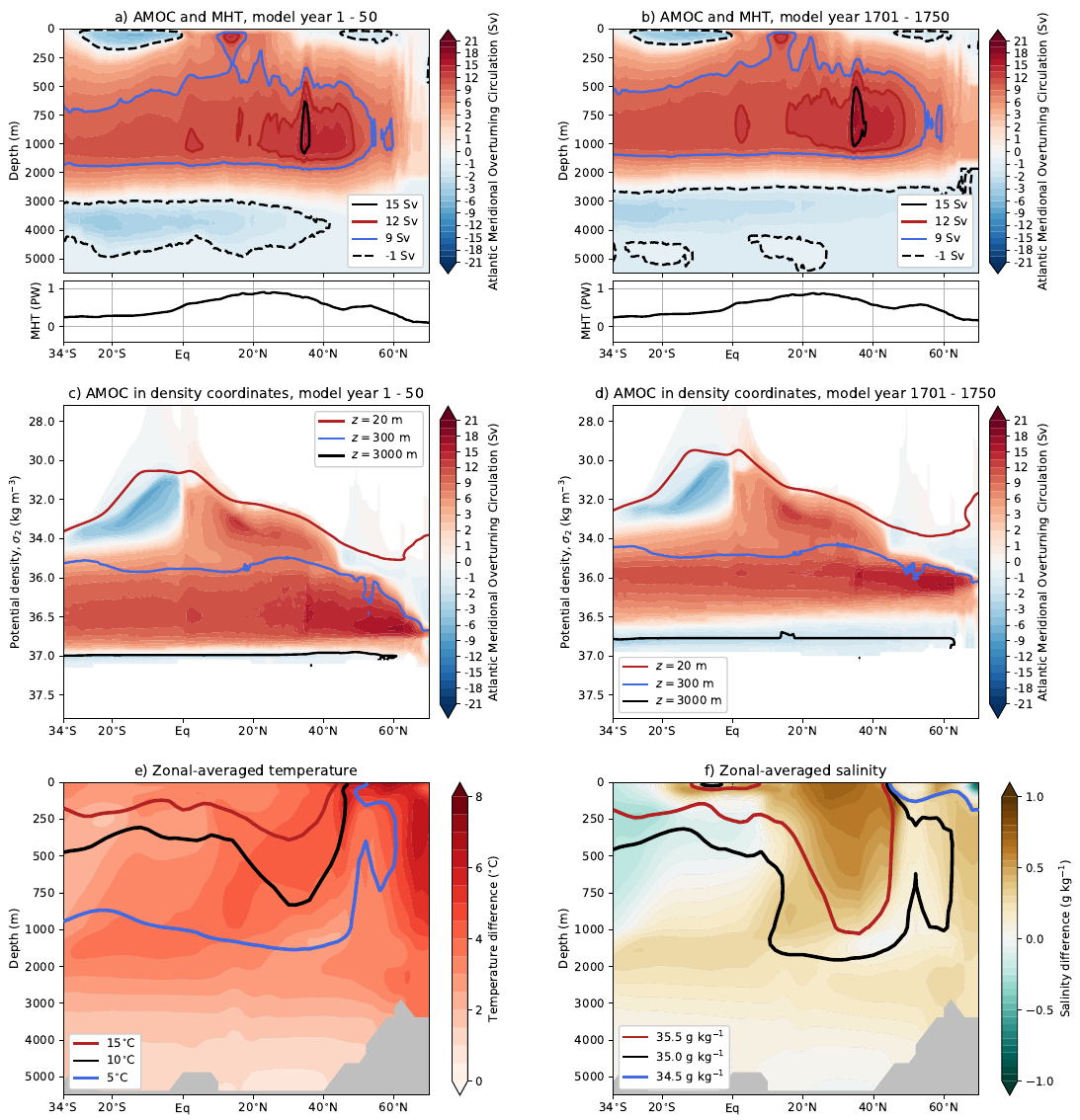} 
\end{tabular}

\caption{\textbf{Oceanic Responses in the Atlantic Ocean} 
(a\&b): The time-mean AMOC in depth coordinates for model years a) 1 -- 50 and b) 1701 -- 1750.
The lower panel shows the meridional heat transport (MHT).
(c\&d): The time-mean AMOC in density coordinates for model years c) 1 -- 50 and d) 1701 -- 1750.
The three curves represent the (section-averaged) depth level, whereas the 20~m depth contour is smoothed to reduce its meridional variability.
(e): The zonally-averaged temperature difference between model years~1701 -- 1750 and PI$_{45}^{\mathrm{on}}$ (shading).
The curves are three isotherms of PI$_{45}^{\mathrm{on}}$.
(f): Similar to panel~e, but now for the zonally-averaged salinity.
The results are from the CO$_2$ ramp of +0.5~ppm~yr$^{-1}$.
}

\label{fig:Figure_S1}

\end{figure}


\begin{figure}[h]

\hspace{-3cm}
\includegraphics[width=1.5\columnwidth, trim = {0cm 0cm 0cm 0cm}, clip]{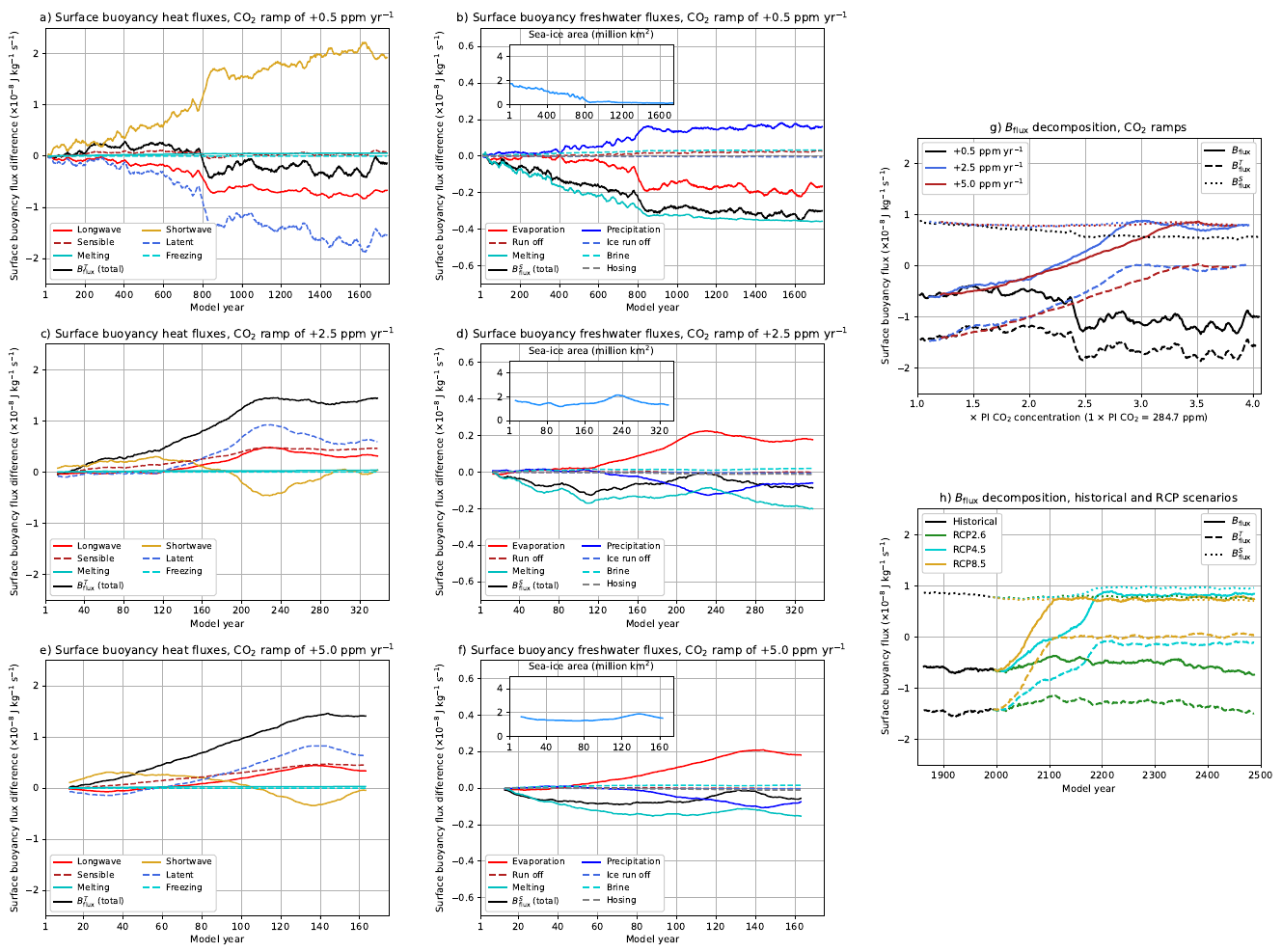}
\caption{\textbf{Surface buoyancy flux decomposition.}
(a -- f): The surface buoyancy flux differences (compared to PI$_{45}^{\mathrm{on}}$) 
between 40$^{\circ}$N -- 65$^{\circ}$N for the three CO$_2$ ramps,
decomposed into the different heat (left column) and freshwater (middle column) fluxes.
The insets shows the yearly-averaged sea-ice area (grid cells with sea-ice fractions of at least 15\%) between 40$^{\circ}$N -- 65$^{\circ}$N.
(g \& h): The $B_{\mathrm{flux}}$ decomposition for the CO$_2$ ramps (in units of PI CO$_2$) and historical and RCP scenarios.
All time series are smoothed through a 25-year running mean to reduce the variability.
}
		
\label{fig:Figure_S2}
\end{figure}


\begin{figure}[h]

\hspace{-3.5cm}
\includegraphics[width=1.5\columnwidth, trim = {0cm 0cm 0cm 0cm}, clip]{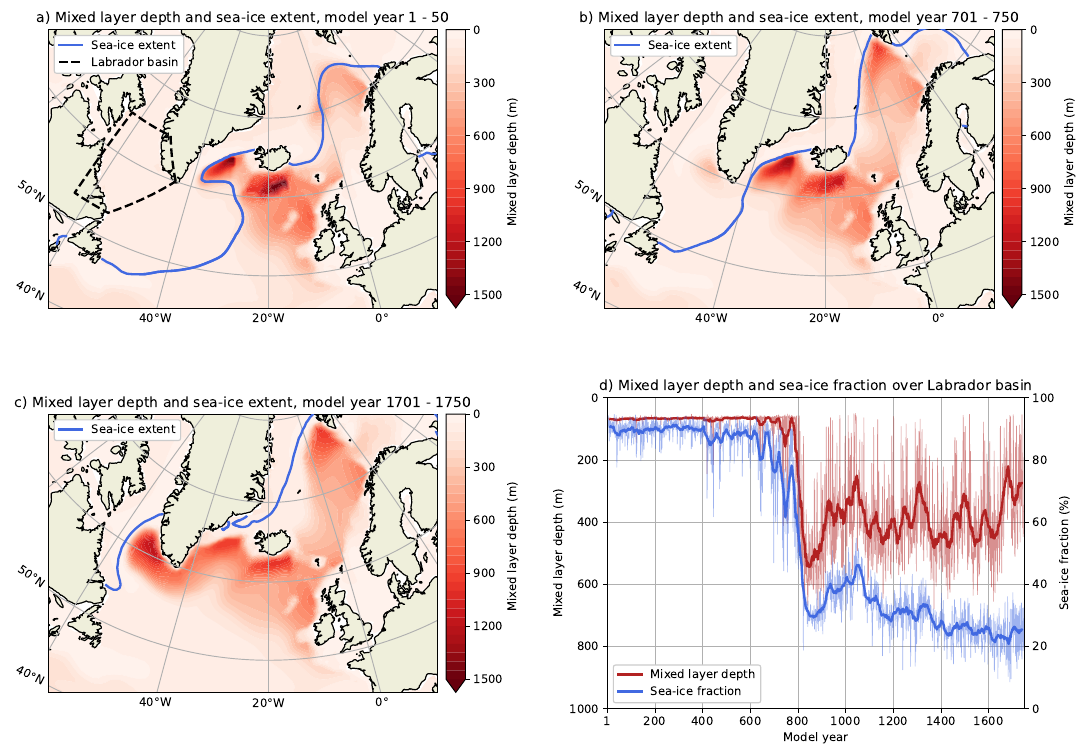}
\caption{\textbf{Mixed layer depth and sea-ice extent.}
(a): The March mixed layer depth and sea-ice extent (i.e., 15\% sea-ice fraction contour) for model years~1 -- 50 of the slow CO$_2$ ramp simulation.
(b \& c): Similar to panel~a, but now for b) model years 701 -- 750 and c) 1701 -- 1750.
(d): The spatially-averaged March mixed layer depth and sea-ice fraction over the Labrador basin (black outlined region in panel~a).
The thin curves are yearly averages, whereas the thick curves are smoothed versions (25-year moving averages).}
	
\label{fig:Figure_S3}
\end{figure}


\begin{figure}[h]

\hspace{-3.5cm}
\includegraphics[width=1.5\columnwidth, trim = {0cm 0cm 0cm 0cm}, clip]{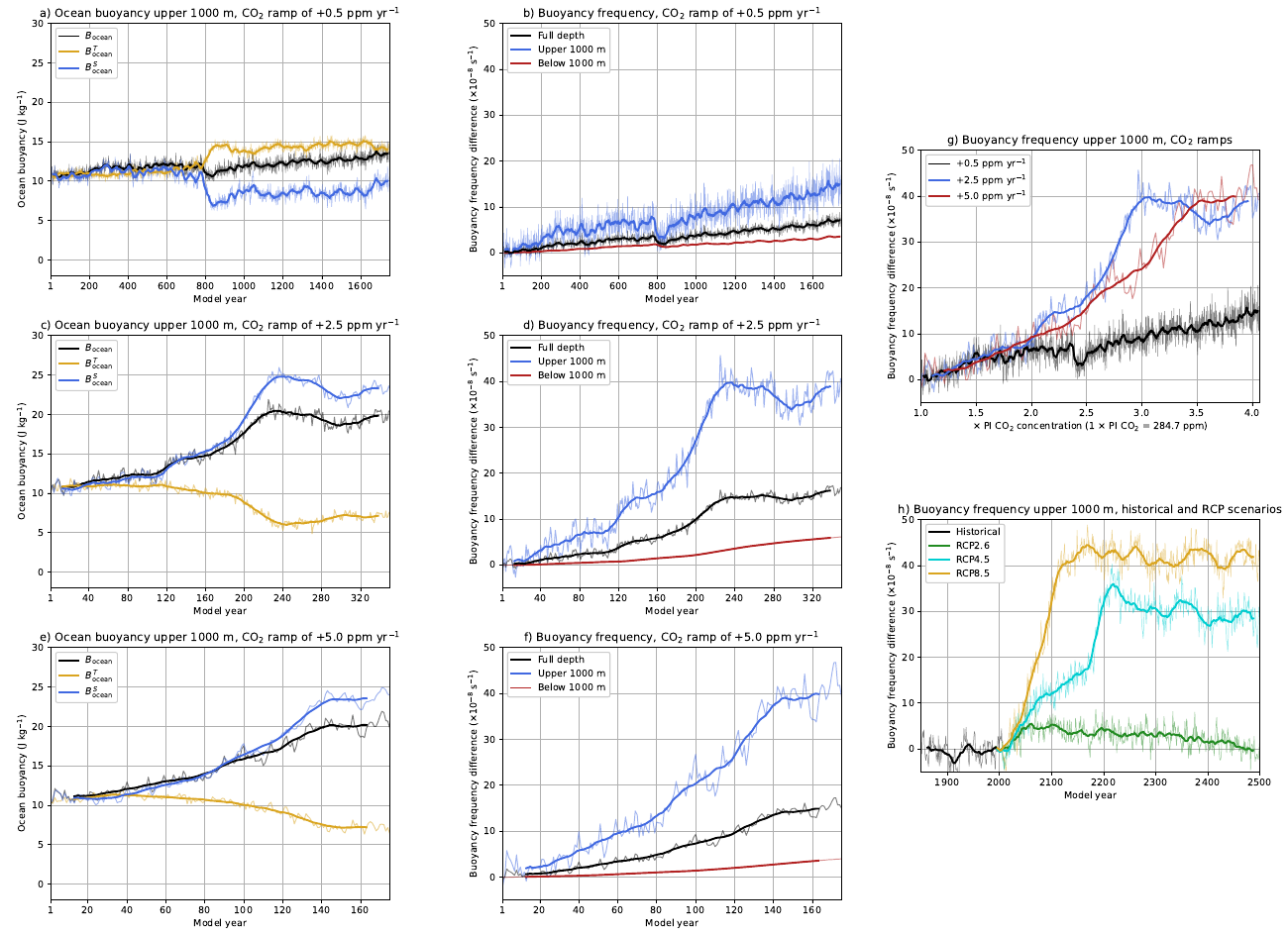}
\caption{\textbf{Upper 1000~m ocean buoyancy.}
(Left column): The upper 1,000~m ocean buoyancy between 40$^{\circ}$N -- 65$^{\circ}$N for the three CO$_2$ ramps.
The ocean buoyancy is decomposed into a temperature and salinity contribution using the fixed salinity and temperature fields from PI$_{45}^{\mathrm{on}}$, respectively.
(Middle column): The buoyancy frequency ($N^2 = \frac{-g}{\rho_0} \frac{\partial \rho}{\partial z}$) difference (compared to PI$_{45}^{\mathrm{on}}$) for the three CO$_2$ ramps,
which is split into an upper 1,000~m and below 1,000~m contribution. 
(Right column): The upper 1,000~m buoyancy frequency difference for the CO$_2$ ramps (compared to PI$_{45}^{\mathrm{on}}$, in units of PI CO$_2$) 
and historical and RCP scenarios (compared to the historical period of 1850 -- 1899).
In all panels, the thin curves are yearly averages, whereas the thick curves are smoothed versions (25-year moving averages).
}
	
\label{fig:Figure_S4}
\end{figure}


\begin{figure}[h]

\begin{tabular}{c}

\includegraphics[width=1\columnwidth, trim = {0cm 0cm 0cm 0cm}, clip]{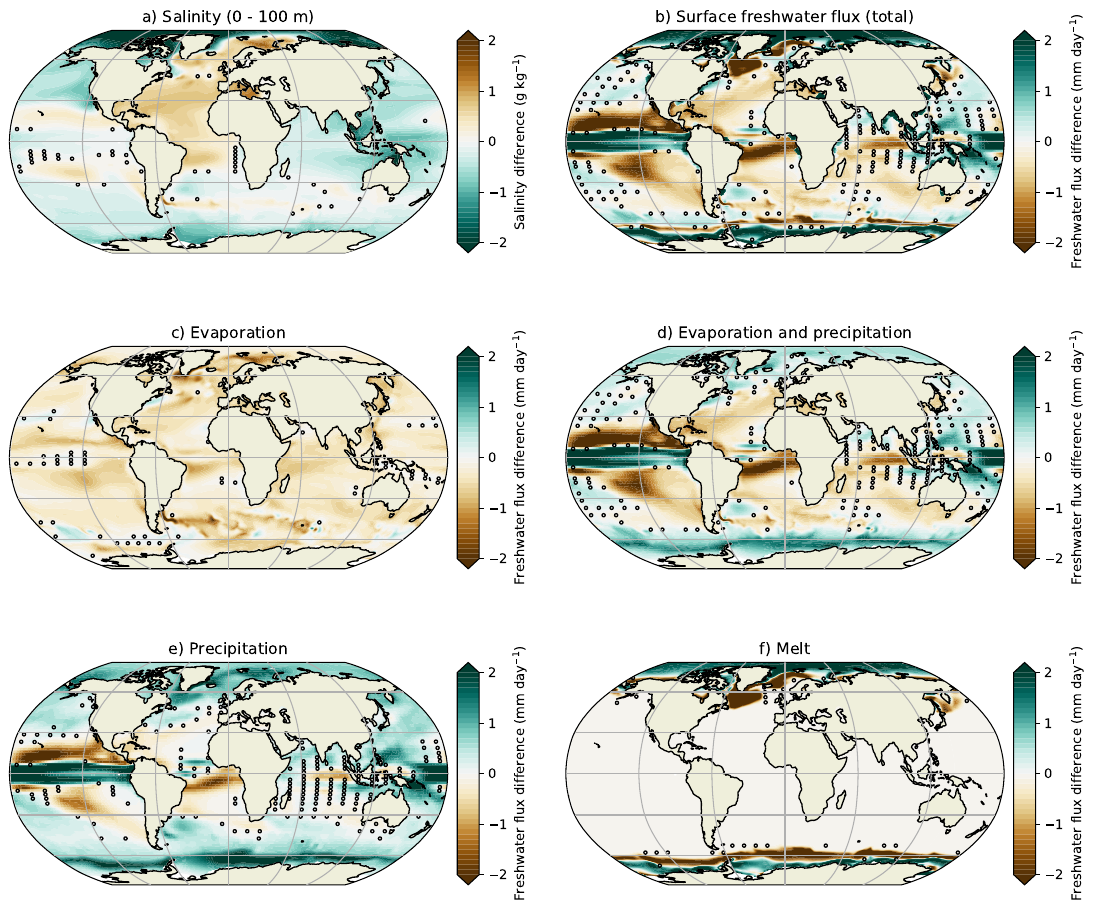} 
\end{tabular}

\caption{\textbf{Salinity and freshwater flux responses} 
(a): The depth-averaged (upper 100~m) salinity differences between model years 1701 -- 1750 of the slow CO$_2$ ramp and PI$_{45}^{\mathrm{on}}$.
(b -- f): The surface freshwater flux differences and its decomposition between model years 1701 -- 1750 of the slow CO$_2$ ramp and PI$_{45}^{\mathrm{on}}$ (positive = relative surface freshening).
In all panels, the circled markers indicate non-significant ($p \geq 0.05$, two-sided Welch t-test) differences,
markers were not displayed when no sea ice was present (panel~f).
}

\label{fig:Figure_S5}

\end{figure}


\begin{figure}[h]

\hspace{-3.5cm}
\includegraphics[width=1.5\columnwidth, trim = {0cm 0cm 0cm 0cm}, clip]{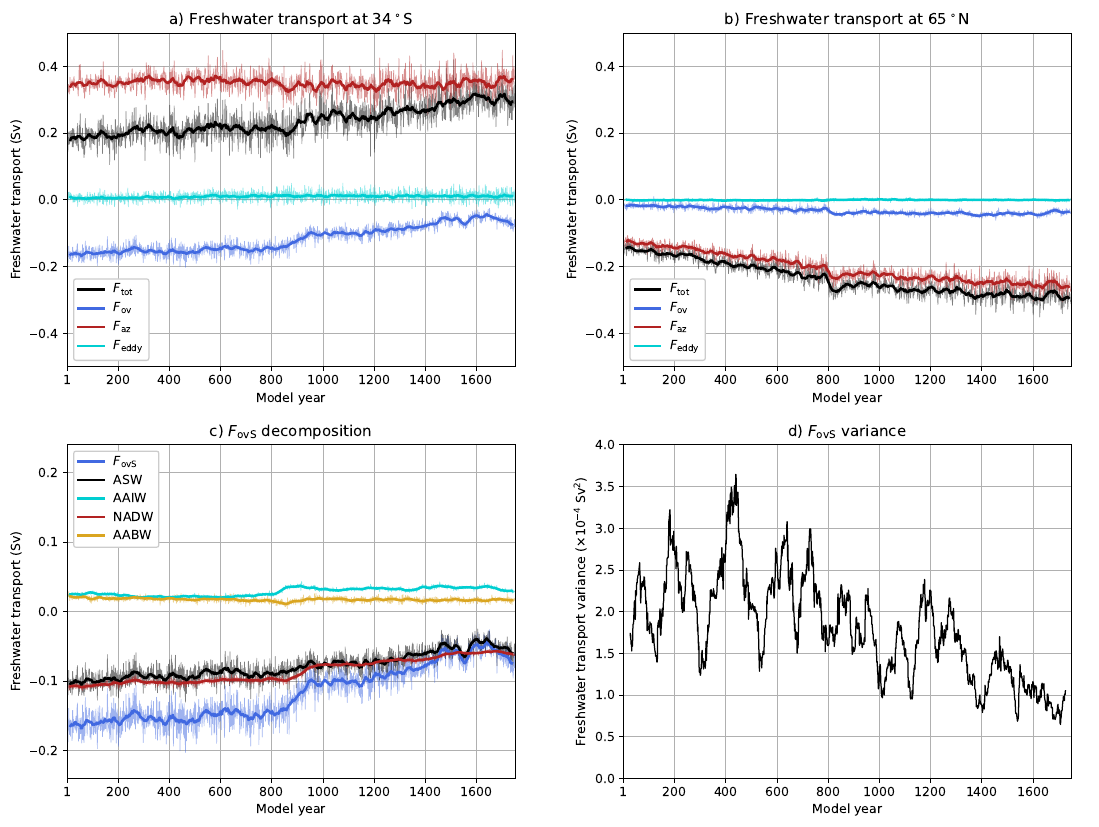}
\caption{\textbf{Freshwater transport decomposition.}
(a \& b): The freshwater transports at 34$^{\circ}$S and 65$^{\circ}$N for the slow CO$_2$ ramp.
(c): The $F_{\mathrm{ovS}}$ decomposition, following the procedure outlined in \cite{vanWesten2024b}.
(d): The $F_{\mathrm{ovS}}$ variance, where the variance is determined over 50-year sliding windows.
A linear trend is removed over the 50-year window before determining the variance.
In panels a,b,c, the thin curves are yearly averages, whereas the thick curves are smoothed versions (25-year moving averages).}
	
\label{fig:Figure_S6}
\end{figure}


\begin{figure}[h]

\hspace{-3.5cm}
\includegraphics[width=1.5\columnwidth, trim = {0cm 0cm 0cm 0cm}, clip]{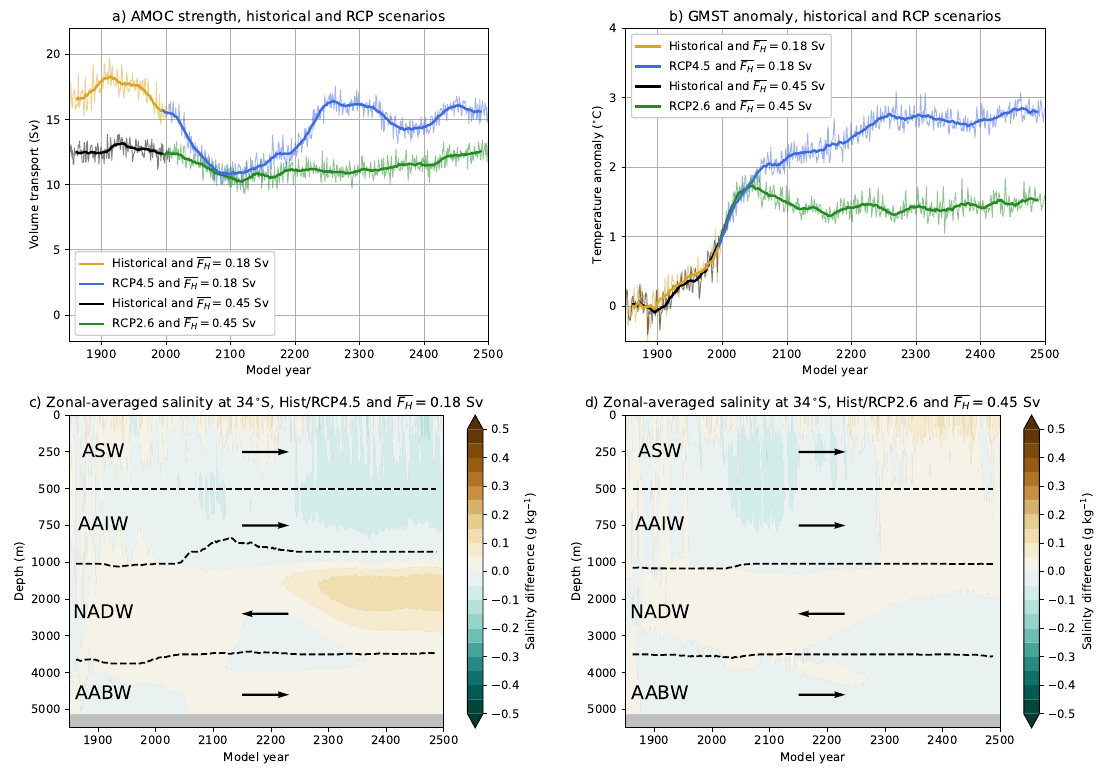}
\caption{\textbf{Climate model simulations with the CESM}
(a): The AMOC strength (at 26$^{\circ}$N and 1,000~m depth) for the historical and RCP4.5 with $\overline{F_H} = 0.18$~Sv, and historical and RCP2.6 with $\overline{F_H} = 0.45$~Sv.
(b): Similar to panel~a, but now for the GMST anomaly compared to 1850 -- 1899.
(c): Similar to Figure~\ref{fig:Figure_5}d, but now for the historical and RCP4.5 with $\overline{F_H} = 0.18$~Sv.
(d): Similar to Figure~\ref{fig:Figure_5}d, but now for the historical and RCP2.6 with $\overline{F_H} = 0.45$~Sv.
In panels a,b, the thin curves are yearly averages, whereas the thick curves are smoothed versions (25-year moving averages).}
	
\label{fig:Figure_S7}
\end{figure}


\begin{figure}[h]

\hspace{-3.5cm}
\includegraphics[width=1.5\columnwidth, trim = {0cm 0cm 0cm 0cm}, clip]{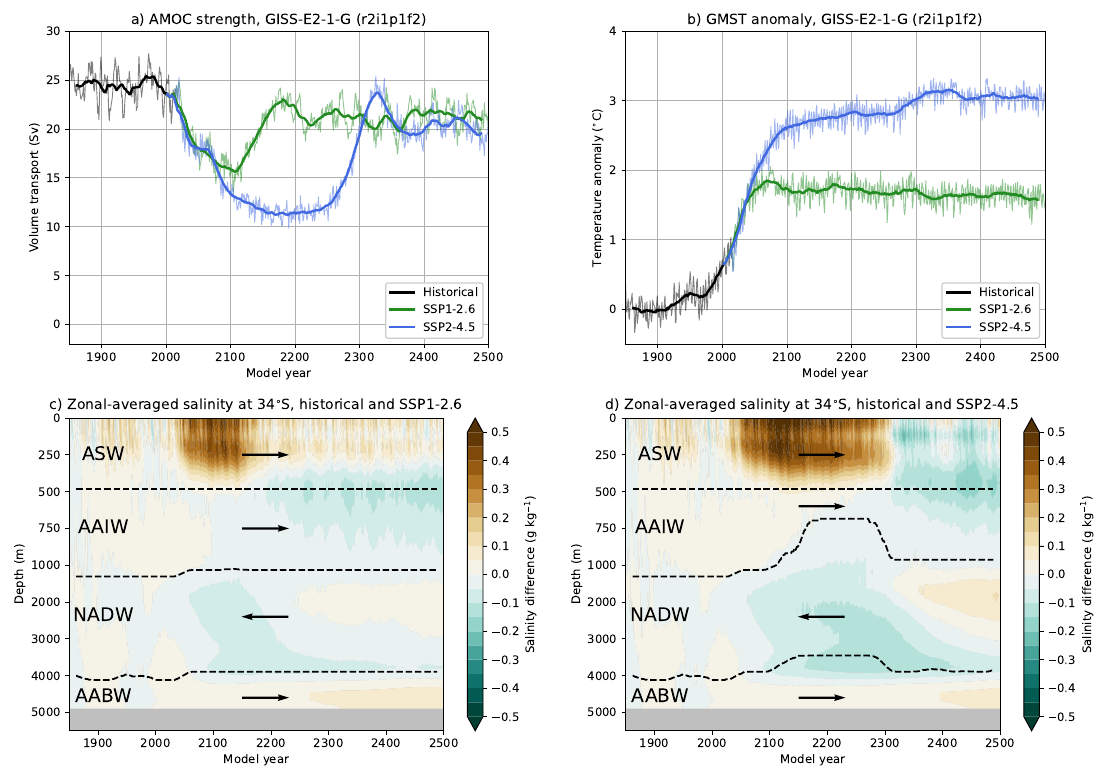}
\caption{\textbf{Climate model simulations with the GISS-E2-1-G (r2i1p1f2)}
Similar to Figure~\ref{fig:Figure_S7}, but now for the GISS-E2-1-G (r2i1p1f2) under the historical and SSP1-2.6 and historical and SSP2-4.5.}
	
\label{fig:Figure_S8}
\end{figure}


\begin{figure}[h]

\hspace{-3.5cm}
\includegraphics[width=1.5\columnwidth, trim = {0cm 0cm 0cm 0cm}, clip]{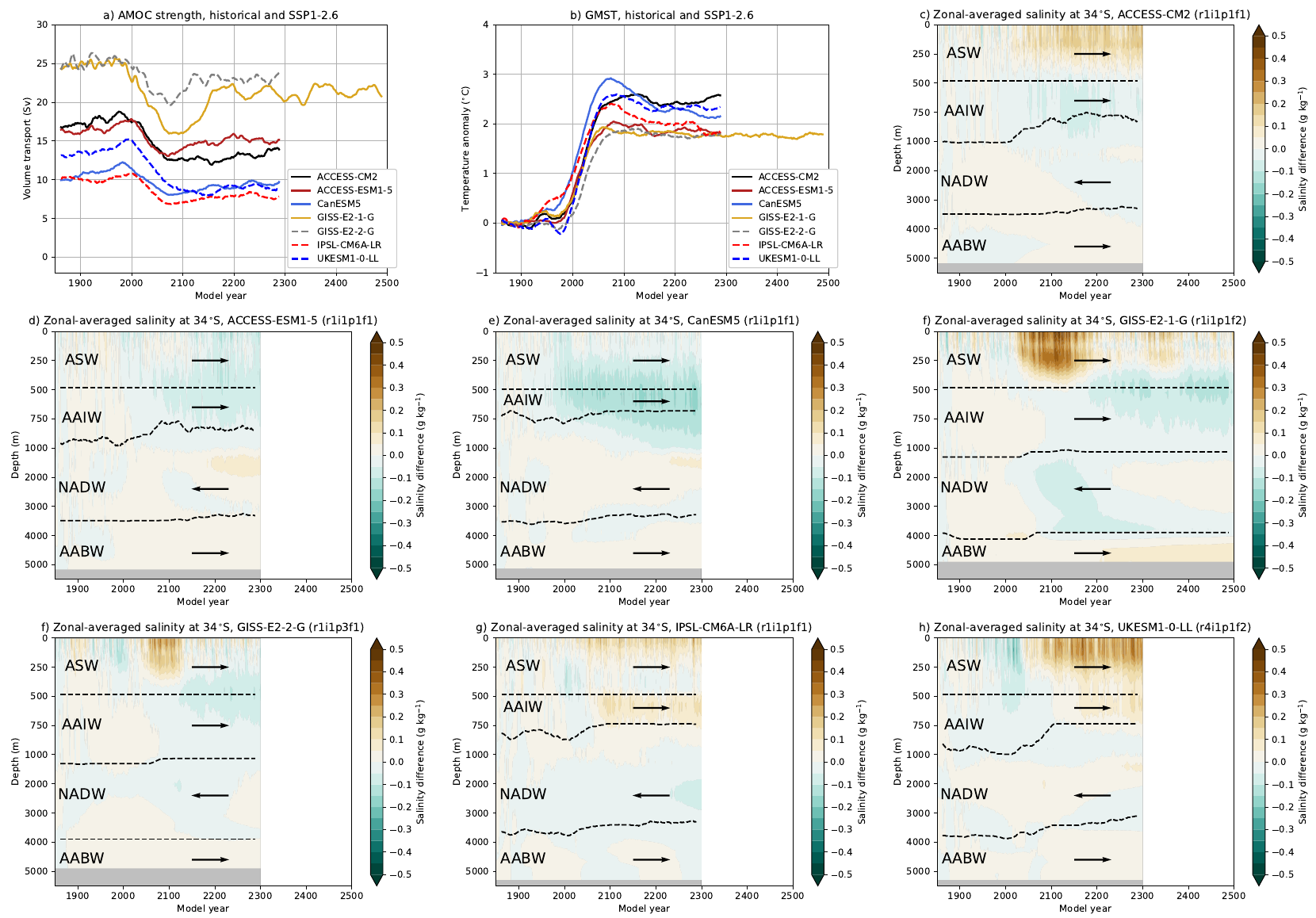}
\caption{\textbf{Climate model simulations under the historical and SSP1-2.6}
Similar to Figure~\ref{fig:Figure_S7}, but now for 7~different CMIP6 models under the historical and SSP1-2.6.
The specific realisations are indicated in the captions of panels~c through h.}
	
\label{fig:Figure_S9}
\end{figure}


\begin{figure}[h]

\hspace{-3.5cm}
\includegraphics[width=1.5\columnwidth, trim = {0cm 0cm 0cm 0cm}, clip]{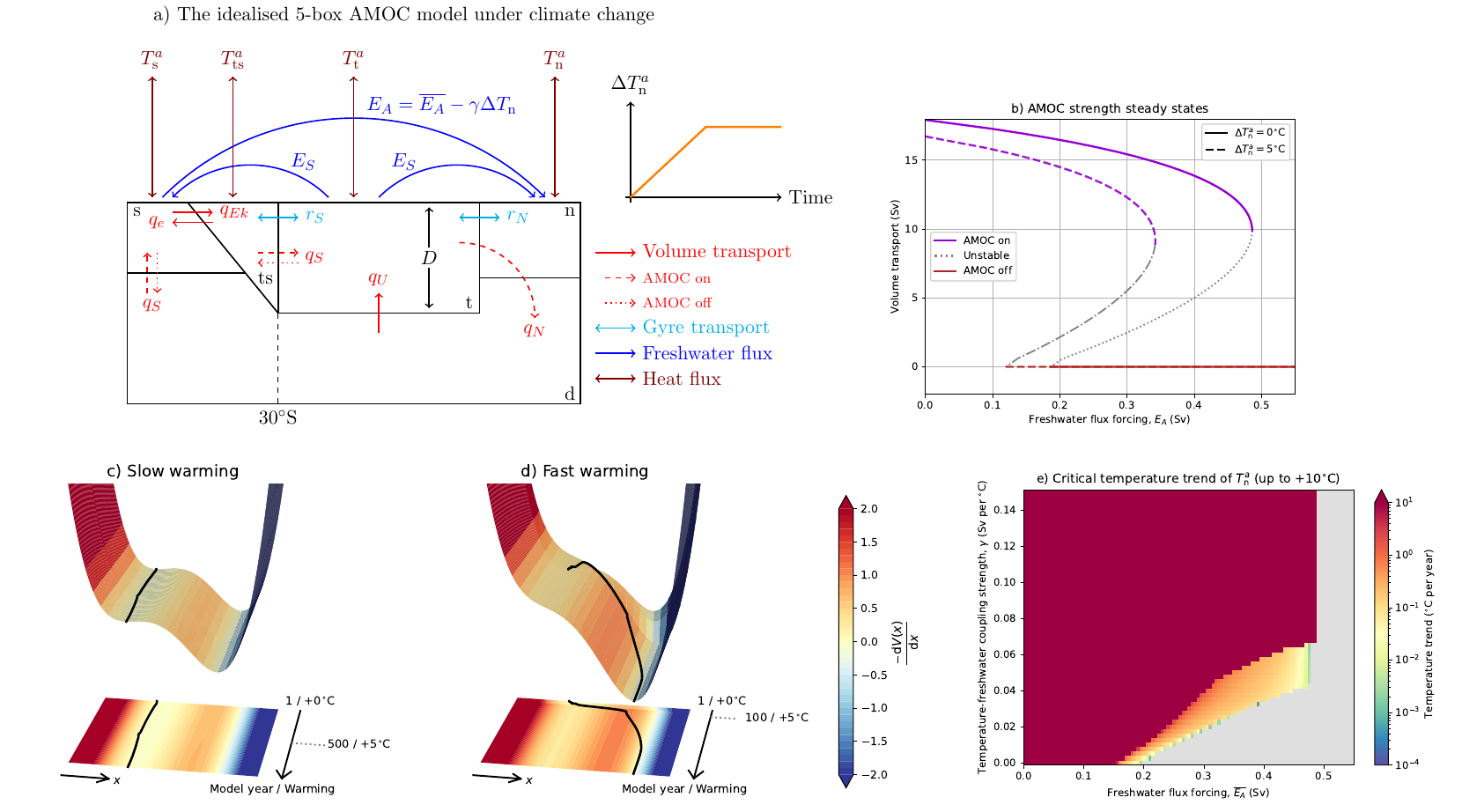}
\caption{\textbf{The  5-box AMOC model}
(a): Schematic of the 5-box AMOC model, with an AMOC on state (clockwise circulation, dashed red arrows) and AMOC off state (anticlockwise circulation, dotted red arrows),
in which temperatures, salinities and pycnocline depth are dynamically resolved.
The model has the asymmetric freshwater flux forcing, $E_A$, as control parameter.
In addition, the model is forced by a varying temperature anomaly $\Delta T_{\mathrm{n}}^a$ applied to $T_\mathrm{n}^a$ (and also $T_{\mathrm{s}}^a$ as they are coupled). 
More details, parameter settings and sensitivity experiments are provided in \cite{vanWesten2024c,vanWesten2025a};
the version analysed here does not consider sea-ice insulation effects.
(b): Bifurcation diagram under varying $E_A$, showing equilibria for fixed $\Delta T_{\mathrm{n}}^a = 0^{\circ}$C and $\Delta T_{\mathrm{n}}^a = 5^{\circ}$C.
(c \& d): Sketch of the (forcing path-dependent) stability landscape as a classical double-well potential,
with the left and right minima representing the `AMOC on' and `AMOC off' states, respectively, shown for the slow and fast warming cases.
In this landscape, the system state is remapped based on the distance to the saddle-node bifurcation, 
where a larger (smaller) distance means that the AMOC is more (less) stable, as is reflected by a relatively deep (shallow) left minimum.
The system trajectory is shown by the black curve. 
(e): Similar to Figure~\ref{fig:Figure_6}d, but now for a linear ramp up to $\Delta T_{\mathrm{n}}^a = 10^{\circ}$C.}
	
\label{fig:Figure_S10}
\end{figure}

\end{document}